\pdfoutput=1
\documentclass[12pt,preprint]{aastex}
\usepackage{amsmath,graphicx,xspace}
\usepackage{enumitem}
\usepackage[colorlinks=true,urlcolor=blue,citecolor=red]{hyperref}
\allowdisplaybreaks
\newcommand{\eg}{{\it e.g.\/}}
\newcommand{\ie}{{\it i.e.\/}}
\newcommand{\etc}{{\it etc.\/}}
\newcommand{\etal}{{\it et al.\/}}
\newcommand{\cf}{{\it c.f.\/}}

\newcommand{\azeus}{\textsf{AZEuS}\xspace}
\newcommand{\zeus}{\textsl{ZEUS}\xspace}
\newcommand{\zeusdd}{\textsl{ZEUS-2D}\xspace}
\newcommand{\zeusddd}{\textsl{ZEUS-3D}\xspace}
\newcommand{\del}{\ensuremath{\partial}}

\newcommand{\cale}{{\cal E}}
\newcommand{\veps}{\varepsilon}
\bibpunct[; ]{(}{)}{;}{a}{,}{,}
\defcitealias{bc89}{BC89}
\defcitealias{rj95}{RJ95}
\defcitealias{t98}{T98}
\shorttitle{AZEuS}
\shortauthors{Ramsey, Clarke \& Men'shchikov}
\slugcomment{Submitted to ApJS on July 6, 2011}
\received{}
\title{\textsf{AZEuS}: An Adaptive Zone Eulerian Scheme for Computational MHD}
\author{Jon P.\ Ramsey\altaffilmark{1}, David A.\ Clarke}
\affil{Institute for Computational Astrophysics, Department of Astronomy \& Physics, Saint Mary's University, Halifax, Nova Scotia B3H 3C3, Canada.}
\and\author{Alexander B. Men'shchikov}
\affil{Laboratoire AIM, CEA/DSM-CNRS-Universit\'e Paris Diderot, IRFU/SAp, CEA Saclay, 91191 Gif-sur-Yvette, France.}
\altaffiltext{1}{Current address: Zentrum f\"ur Astronomie der Universit\"at Heidelberg, Institut f\"ur Theoretische Astrophysik, Albert-Ueberle-Str. 2, 69120 Heidelberg, Germany.}
\begin{abstract}
A new adaptive mesh refinement (AMR) version of the \zeusddd astrophysical magnetohydrodynamical (MHD) fluid code, \azeus, is described.  The AMR module in \azeus has been completely adapted to the staggered mesh that characterises the \zeus family of codes, on which scalar quantities are zone-centred and vector components are face-centred.  In addition, for applications using static grids, it is necessary to use higher-order interpolations for prolongation to minimise the errors caused by waves crossing from a grid of one resolution to another.  Finally, solutions to test problems in 1-, 2-, and 3-dimensions in both Cartesian and spherical coordinates are presented.
\end{abstract}
\keywords{hydrodynamics --- magnetohydrodynamics (MHD) --- methods: numerical}
\begin{document}
\maketitle
\section{Introduction}
\label{sec:intro}
High-resolution, multidimensional simulations have become indispensable for many complex problems in astrophysics, particularly those involving (magneto-)fluid dynamics.  One of the most important innovations in this area has been the use of dynamic and variable resolution techniques.  Adaptive mesh refinement (AMR), pioneered in the context of the fluid equations by \citet{bo84} and \citet[\citetalias{bc89}]{bc89}, is one such approach.

With AMR, a hierarchy of grids is used to provide high numerical resolution when and where the physics requires it, leaving as much of the volume at lower resolution as possible to minimise computational effort.  This makes AMR an efficient means of studying problems with a very large spatial dynamic range (\eg, star formation, galaxy evolution), as borne out by the large number of codes which employ it: \textsl{ORION} \citep{k99}, \textsl{FLASH} \citep{f00}, \textsl{RIEMANN} \citep{b01}, \textsl{RAMSES} \citep{fht06}, \textsl{PLUTO} \citep{m07}, \textsl{NIRVANA} \citep{z08}, \textsl{AstroBEAR} \citep{cfvmj09}, and \textsl{ENZO} \citep{cxnll10} to name several.

Virtually all AMR fluid codes to date are based on a zone-centred grid, with all hydrodynamical variables (density, energy, and momentum components) taken to be located at the centres of their respective zones.  Indeed, AMR was originally designed specifically for zone-centred schemes.  Magnetohydrodynamic (MHD) solvers are designed with either zone-centred or face-centred magnetic field components, depending in part on the mechanism used to preserve the solenoidal condition.  One scheme that has enjoyed somewhat of a renaissance of late is Constrained Transport (CT; \citealt{eh88}), which places magnetic field components at the centres of the zone-faces to which they are normal.  The staggered mesh introduced in such a scheme has to be specifically accounted for in the AMR modules and in such a way that $\nabla \cdot \vec B$ remains zero everywhere---including within the boundaries---to machine round-off error.

\begin{figure}[t!]
  \begin{center}
    \includegraphics[scale=0.5]{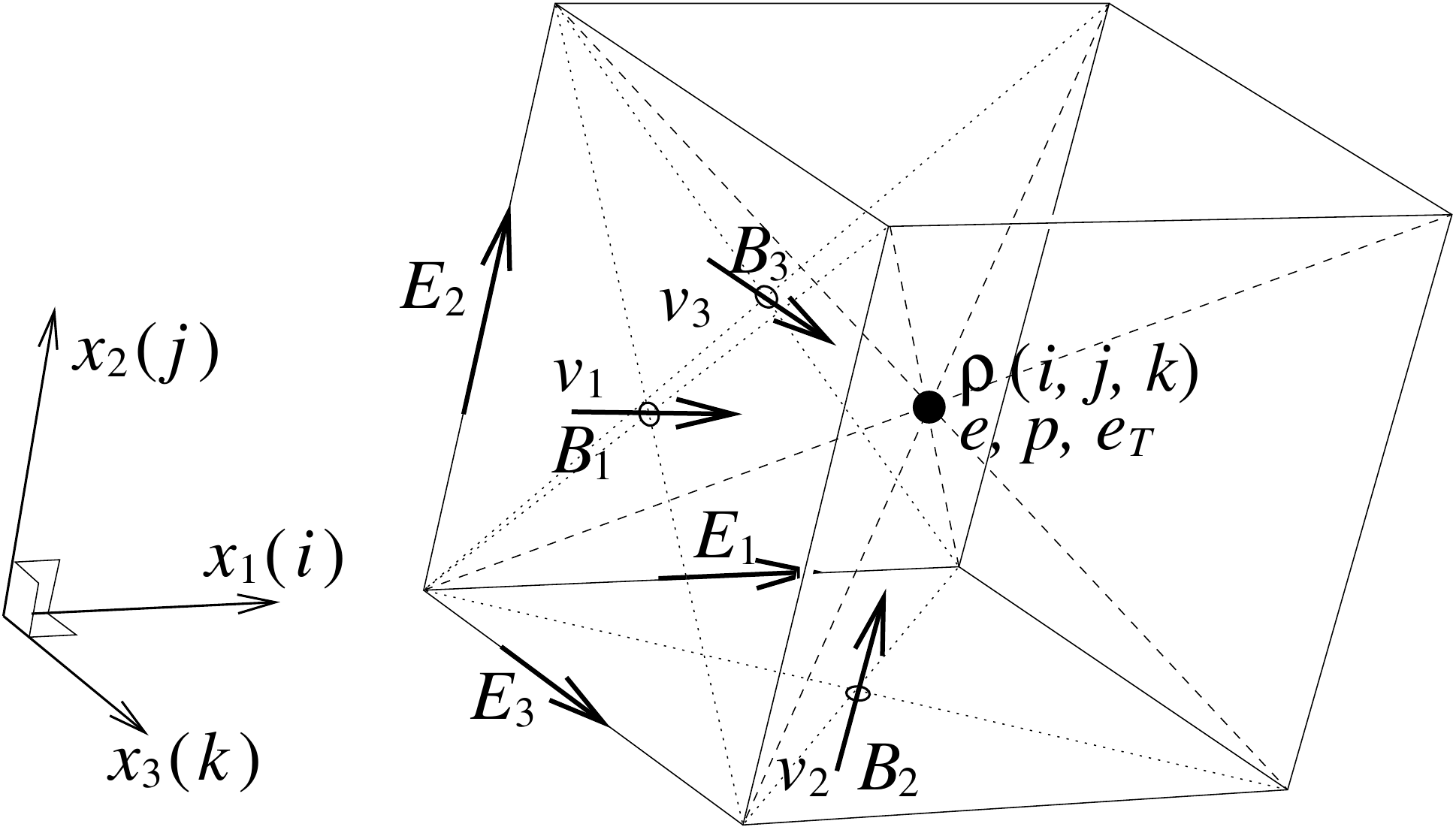}
  \end{center}
  \caption{\label{fig:staggrid} Depiction of a single zone on a fully-staggered grid, where scalars ($\rho$, $e_{\rm T}$, $e$, $p$) are zone-centred, primitive vectors ($\vec{v}$, $\vec{B}$) are face-centred, and derived vectors ($\vec{E} = - \vec{v} \times\vec{B}$, $\vec J = \nabla \times \vec B$) are edge-centred.}
\end{figure}

The \zeus family of codes are one of only a very few astrophysical fluid codes in use that employ a \emph{fully} staggered grid (\eg, \textsl{STAGGER}; \citealt{kritsuketal2011} and references therein), where the momentum components are also face-centred (Figure \ref{fig:staggrid}).  While concerns have been expressed over the suitability of its MHD algorithms in certain pedagogical 1-D test problems (\eg, \citealt{f02}), the fact remains that in one form or another, \zeus is among the best tested, documented, and widely-used fluid codes in astrophysics (\citealt{sn92a,sn92b,smn92,c96,c07,c10,hayesetal06}), and a proper merger with AMR is warranted.  To do this, AMR has to be modified for a fully staggered grid, including the proper treatment of face-centred magnetic field and face-centred momentum.

In this paper, we introduce the newest member of the \zeus family of codes, \azeus, whose ``maiden simulations" have already appeared in \citet{rc11}.  \azeus is a block-structured AMR version of \zeusddd \citep{c96,c10} which preserves the modularity and structure of the underlying \zeus module.  The AMR scheme of \citetalias{bc89}, including the changes described in \citet{bbsw94}, are modified for a fully-staggered grid with additional modifications made to the prolongation procedure to allow for smooth passage of all types of waves between adjacent grids of differing resolution.  \azeus is currently capable of ideal MHD in 1-, 1.5-, 2-, 2.5-, and 3-D in Cartesian, cylindrical, and spherical polar coordinates using both dynamic and static grids, and with a full suite of physical boundary conditions.  As with all \zeus-type codes, its operator-split design allows for additional physics (\eg, gravity, viscosity, radiation, \etc) to be added without concern over how such additions will affect the MHD algorithm.  How the non-hyperbolic additions are implemented for an adaptive grid is another matter (\eg, radiation; \citealt{wa11}).

This paper does not attempt to give a full recount of the basic methodology in either AMR or \zeus, but focuses instead on the modifications to AMR (not so much to \zeus) necessary for their merger.  Thus, the reader should be familiar with \citetalias{bc89} and \citet{c96,c10}.  In Section \ref{sec:preamble}, we list the MHD equations solved by \azeus, and define our conventions and notation.  In Sections \ref{sec:restrict} and \ref{sec:prolong}, we enumerate the modifications necessary for restriction and prolongation on a staggered grid, as well as outline the interpolation schemes used to allow the smooth passage of waves across grid boundaries.  Section \ref{sec:bcs} focuses on boundary conditions, while in Section \ref{sec:gcreate} we discuss how grids are created and how the proper nesting criterion must be modified for a fully-staggered grid.  Several of the 1-D, 2-D, and 3-D test problems used to validate \azeus are given in Section \ref{sec:tests}, followed by a quick summary in Section \ref{sec:summary}.  Discussion of curvilinear coordinates, use of the vector potential, and a schematic overview of the code are relegated to the appendices.
\section{Preamble}
\label{sec:preamble}
\subsection{Underlying numerical method}
\label{sub:underlying}
\azeus solves the following equations of ideal MHD (with the artificial viscosity and gravity terms included):
\begin{align}
  \label{eq:continuity}
  \frac{\del\rho}{\del t} + \nabla\cdot\left(\rho\vec{v}\right) & ~=~ 0; \\
  \label{eq:momentum}
  \frac{\del\vec{s}}{\del t} + \nabla\cdot\left(\vec{s}\,\vec{v}\right) & ~=\,\, -\nabla p + (\nabla\times\vec B)\times \vec B -\nabla\cdot \mathsf{Q} - \rho\nabla \phi; \\
  \label{eq:induction}
 \frac{\del\vec{B}}{\del t} + \nabla \times \vec E & ~=~ 0; \\
  \label{eq:intnrg}
  \frac{\del e}{\del t} + \nabla\cdot( e\vec{v}) & ~=\,\, -p\nabla\cdot\vec{v} - \mathsf{Q} : \nabla\vec{v}; \\
  \label{eq:totnrg}
  \frac{\del e_{\rm T}}{\del t} + \nabla\cdot\big(e_{\rm H}\vec{v} + \vec{E}\times\vec{B} + \mathsf{Q}\cdot \vec v\big) & ~=~ 0,
\end{align}
where $\rho$ is the mass density, $\vec{v}$ is the velocity, $\vec{s} = \rho\vec{v}$ is the momentum density, $p$ is the thermal pressure, $\vec{B}$ is the magnetic induction\footnote{in units where $\mu_0 = 1$}, $\mathsf{Q}$ is the von Neumann-Richtmyer artificial viscous stress tensor \citep{vNR50,c10}, $\phi$ is the gravitational potential and satisfies the Poisson equation ($\nabla^2\phi=4\pi G\rho$), $\vec{E} = - \vec{v}\times\vec{B}$ is the induced electric field, $e$ is the internal energy density, $e_{\rm H} = e + \onehalf\rho v^2 + \rho\phi$ is the hydrodynamical energy density, and $e_{\rm T} = e_{\rm H} + \onehalf B^2$ is the total energy density.  This set of equations [in which equation (\ref{eq:totnrg}) can be derived from equations (\ref{eq:continuity})--(\ref{eq:intnrg})] is closed by the ideal gas law, $p = (\gamma - 1)e$, where $\gamma$ is the ratio of specific heats.  Figure \ref{fig:staggrid} shows the locations of most of these variables on a fully-staggered grid.  Other physics terms often found in \zeus codes such as a second fluid, physical viscosity, radiation, \etc, have yet to be implemented.

\azeus inherits the operator-split methodology of \zeusddd, wherein the terms on the RHS of equations (\ref{eq:continuity})--(\ref{eq:totnrg}) are treated in a \emph{source step} while those on the LHS are accounted for in a separate \emph{transport} and \emph{inductive step}.  As such, the algorithm is not strictly conservative.  However, based as it is on the version of \zeusddd described by \citet{c10}, \azeus can solve either the internal energy equation or the total energy equation, where the latter choice does ensure conservation of total energy to machine round-off error, but at the cost of non-positive-definite thermal pressure.  Should positive definite pressures be paramount, the internal energy equation offers a viable option with, in most cases, total energy conserved to within 1\% or less (see \citealt{c10} for further discussion).

\begin{figure}[t!]
  \begin{center}
    \includegraphics[scale=0.5]{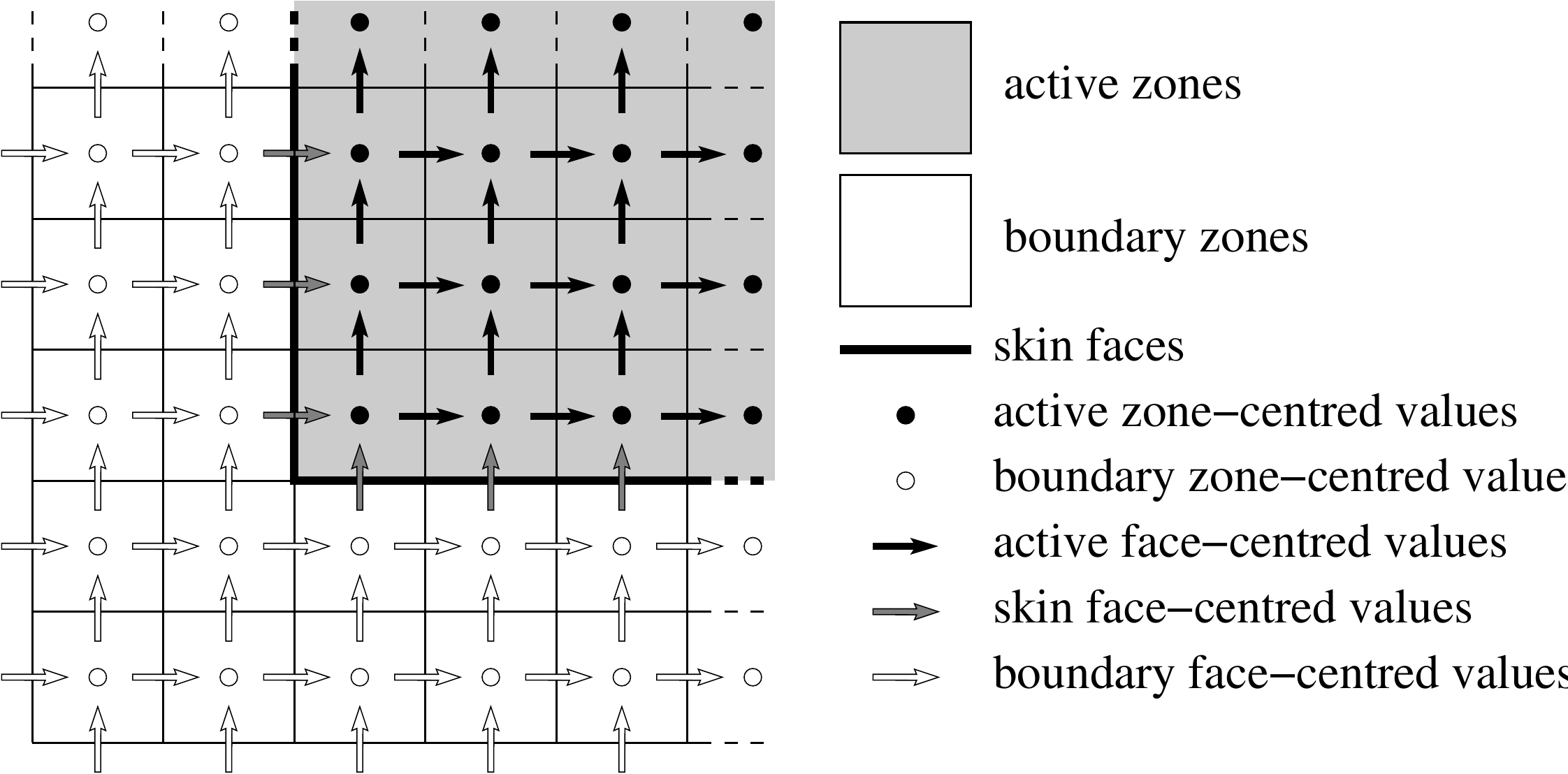}
  \end{center}
  \caption{\label{fig:boundaries} On a fully-staggered grid, all variables have two boundary values.  In addition, for each direction, one component of a face-centred vector and two components of an edge-centred vector have one skin value.}
\end{figure}

To accommodate the interpolation schemes, two boundary zones must be specified at the edges of all grids.  On a staggered grid, all zone-centred quantities have just the two boundary values, while face-centred quantities have two boundary values plus a value that lies on the face separating the ``active zones" from the ``boundary zones", henceforth referred to as the ``skin" of the grid (Figure \ref{fig:boundaries}).  As we shall see, skin values for the magnetic field are treated just like active zones, while the momenta on the skin are treated somewhere in between active and boundary values; the difference attributed to the conservation properties of these two quantities.

\subsection{Conventions and Notation}
\label{sub:notation}
\begin{figure}[t!]
  \begin{center}
    \includegraphics[width=0.9\textwidth]{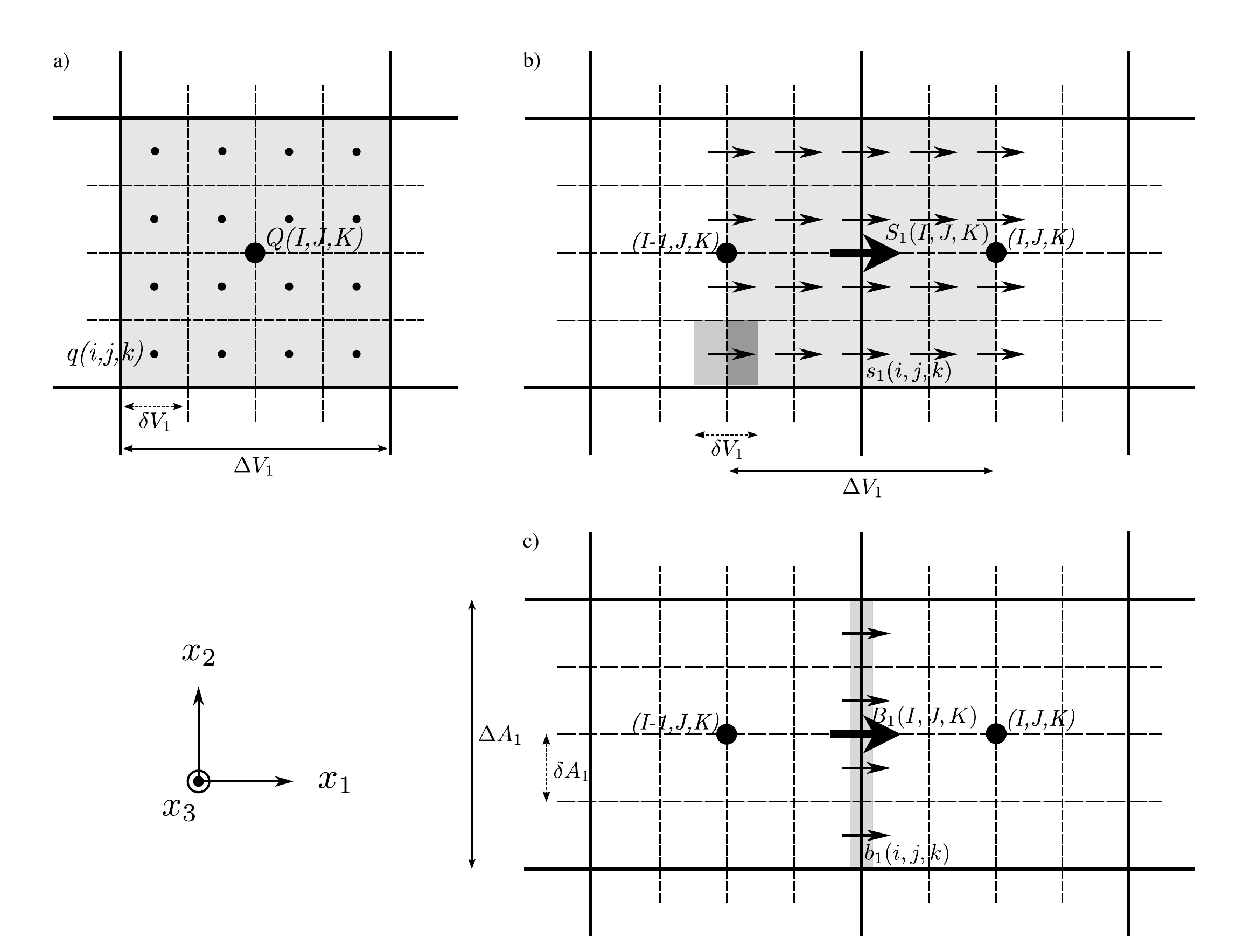}
  \end{center}
  \caption{\label{fig:rois} The regions of influence (ROI) (shaded) for: (a) zone-centred variables, (b) face-centred and volume-conserved variables, and (c) face-centred and area-conserved variables.  A refinement ratio of $\nu=4$ is shown.}
\end{figure}

We adopt the following conventions and notation throughout this paper:

\begin{enumerate}
  \item Quantities in coarse and fine zones are denoted with upper and lower cases respectively: \eg, $Q(I,J,K)$, $q(i,j,k)$.  Fluxes for a quantity $Q$ ($q$) are denoted $F_{m,Q}$ ($f_{m,q}$), where $m =$ 1, 2, or 3 indicates the component direction.
  \item If a ``coarse'' grid or zone is considered to be at level $l$, its daughter ``fine'' grid or zone is at level $l+1$.  The base and coarsest grid, which covers the entire domain, is at level $l = 1$.  The refinement ratio, $\nu$, between level $l$ and $l+1$ must be a power of 2 and the same in all directions.
  \item Grid volumes, areas, lengths and time steps are $\Delta V$, $\Delta A_m$, $\Delta x_m$, and $\Delta t$ for a coarse grid, and $\delta V$, $\delta A_m$, $\delta x_m$, and $\delta t$ for a fine grid.
  \item Indices $(i,j,k)$ correspond to the fine zone at the (left, bottom, back) of a coarse zone with indices $(I,J,K)$.  The zone-centre of a particular fine zone within a coarse zone is designated $(i+\alpha,j+\beta,k+\eta)$, where $\alpha, \beta, \eta = 0,1, \dots\, \nu-1$.  For the 1-face-centre of a fine zone, $\alpha= 0,1, \dots\, \nu$ while $\beta, \eta = 0,1, \dots\, \nu-1$, \etc
  \item Similarly, grid positions for the coarse grid use upper case indices [\eg, $x_1(I)$], while grid positions for the fine grid use lower-case indices [\eg, $x_1(i)$].
  \item The current time step for the coarse grid is indicated by the upper case superscript $N$, while the current fine time step is indicated by the lower case superscript $n$.  Typically, $n=\nu N$.  To designate a fine time step within a coarse time step, we use $n + \tau$ where $0\leq \tau \leq \nu-1$\footnotemark.

\footnotetext{While a coarse zone is always divided in each direction into $\nu$ fine zones, a coarse time step isn't necessarily divided into $\nu$ fine time steps; additional fine time steps are taken if local CFL conditions demand it.  For simplicity of description, however, we shall proceed as though $n = \nu N$, though the reader should be aware that this may not always be true.}

  \item The \emph{region of influence} (ROI) of a variable is defined as the area or volume over which that quantity is conserved.  For zone-centred scalars on the coarse grid, this is the volume $\Delta V (I,J,K)$ (Figure \ref{fig:rois}a).  For face-centred but volume-conserved quantities (\eg, $S_1$), the ROI is the \emph{staggered} volume $\big[$\eg, $\frac{1}{2}\big(\Delta V (I,J,K) + \Delta V (I-1,J,K)\big)\big]$ (Figure \ref{fig:rois}b).  Finally, for face-centred but area-conserved quantities (\eg, $B_1$), the ROI is the area of the face at which the vector component is centred [\eg, $\Delta A_1(I,J,K)$] (Figure \ref{fig:rois}c).
\end{enumerate}

Finally, while \azeus is written in the covariant fashion of \zeusdd \citep{sn92a}, our discussion is given in terms of Cartesian-like components with uniform zone sizes within each grid for simplicity.  As such, $\Delta V/\delta V = \nu^3, \Delta A_m/\delta A_m = \nu^2, \Delta x_m/\delta x_m = \nu$, and $\Delta t/\delta t = \nu$.  Some of the modifications necessary for curvilinear coordinates are given in Appendix \ref{app:curvi}.

\section{Restriction}
\label{sec:restrict}
Restriction is the process by which data on the coarse grid are replaced by an average of data from an overlying fine grid.  This must be done in a fashion that locally preserves all conservation laws and the solenoidal condition to within machine round-off error.  Two types of restriction are considered: the conservative overwrite of coarse values with ROIs which are entirely covered by ROIs of overlying fine zones, and flux corrections to coarse zones (sometimes called ``refluxing'') with ROIs which are adjacent to, or partially covered by, the ROIs of fine zones.
\subsection{Conservative Overwrite}
\label{sub:overwrite}
At the end of a coarse time step, fine and coarse grids are synchronised by overwriting the coarse grid with ``better'' values from overlying fine grids.  Because of the different ROIs on a staggered mesh, the specifics of the overwriting procedure depend on which variable is being overwritten.  For zone-centred, volume-conserved quantities (\eg, $\rho$, $e$, $e_{\rm T}$), the procedure is the same as in \citetalias{bc89}:
\begin{equation}
  \label{eq:over_zcvc}
  Q(I,J,K) ~=~ \frac{1}{\nu^3} \sum_{\alpha, \beta, \eta = 0}^{\nu-1}\, q(i+\alpha, j+\beta, k+\eta),
\end{equation}
where the sum is a triple sum.  By inspection, one can confirm that equation  (\ref{eq:over_zcvc}) conserves $Q$ ($q$) locally to within machine round-off error.

For face-centred, volume-conserved quantities such as the momentum density whose ROIs are completely covered by the ROIs of overlying fine zones, we have:
\begin{equation}
  \label{eq:over_fcvc}
  S_1(I,J,K) ~=~ \frac{1}{\nu^3} \sum_{\alpha=-\nu/2}^{\nu/2}\, \sum_{\beta, \eta = 0}^{\nu-1}\, {\cal G}(\alpha)\, s_1(i+\alpha, j+\beta, k+\eta),
\end{equation}
\begin{equation}
  \label{eq:gfactor}
  \textrm{where~~} {\cal G}(\alpha) ~=~
  \begin{cases}
    1/2 & \textrm{if } \alpha = \pm\, \nu/2 \\
    1 & \textrm{otherwise},
  \end{cases}
\end{equation}
for the 1-component of the momentum.  The factor ${\cal G}(\alpha)$ takes into account that only half of the ROIs of the fine momenta at $\alpha = -\nu/2$, $\nu/2$ cover the ROI of coarse momentum (Figure \ref{fig:rois}b).  

Coarse momenta with ROIs partially covered by a fine grid are co-spatial with the skin of the overlying fine grid.  As skin values of momenta are considered to be boundary values (since one of the fluxes is completely determined from within the boundary), they are not taken to be more reliable than the underlying coarse values (whose fluxes are determined exclusively by interior zones), and thus the coarse values are not overwritten by the fine grid values.  Instead, coarse momenta cospatial with a fine grid skin are considered to be adjacent to the fine grid and, as such, are subject to the ``flux-correction" step described in the next subsection.

For magnetic field, the conserved quantity is the magnetic flux ($\int \vec{B}\cdot d\vec{A}$).  Thus, coarse values of $B_1$ are overwritten using:
\begin{equation}
  \label{eq:over_fcac}
  B_1(I,J,K) ~=~ \frac{1}{\nu^2} \sum_{\beta, \eta ~=~ 0}^{\nu-1}\, b_1(i, j+\beta, k+\eta),
\end{equation}
where the sum is over all fine ROIs (areas of 1-faces) that cover the coarse ROI.  One can easily show that overwritten values of $\vec{B}$ will still satisfy the solenoidal condition---even when combined with values of $\vec B$ that are not overwritten---so long as the overlying values of $\vec{b}$ are divergence-free and adjacent values of $\vec B$ are properly ``refluxed" (Section \ref{sub:reflux}).  In addition, since $\vec{B}$ is an area-conserved quantity, there is never partial coverage of ROIs as there is with the momenta, and thus no analogue of ${\cal G}(\alpha)$ (equation \ref{eq:gfactor}) is necessary for the magnetic field.

A straightforward permutation of indices gives the analogous expressions for the other components of momentum and magnetic field.
\subsection{Flux Corrections}
\label{sub:reflux}
A coarse zone adjacent to but not covered by a fine grid shares a face with the fine grid.  In order that local mass, momentum, and magnetic flux remain conserved to within machine round-off error, the coarse and fine zones must agree on the fluxes passing across their common face.  This is accomplished by keeping track of all coarse fluxes passing across the skin of a contained fine grid, and then subtracting from these the fine fluxes computed during the MHD updates of the fine grids.  These ``flux corrections" are then subtracted from the coarse zones adjacent to the fine grid during the so-called ``refluxing step", effectively replacing the coarse fluxes with the fine fluxes along their common face.

\begin{figure}[t!]
  \begin{center}
    \includegraphics[width=0.9\textwidth]{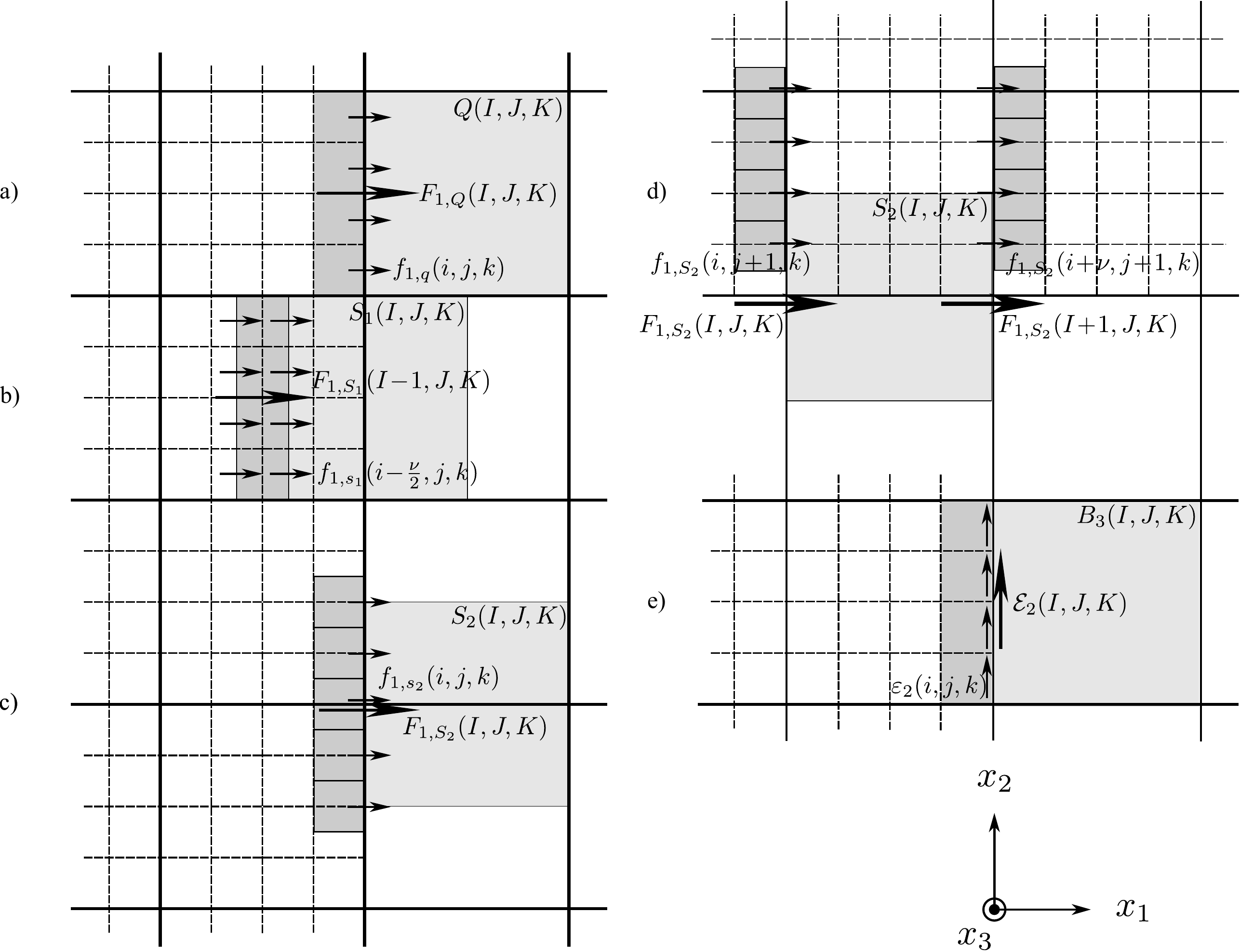}
  \end{center}
  \caption{\label{fig:reflux} The different cases for flux corrections on a staggered grid, including: (a) zone-centred quantities; (b, c, and d) the three different cases for face-centred, volume-conserved momenta; and (e) area-conserved magnetic field corrections via the EMFs.  Shaded regions denote typical ROIs referred to in the text.  Note that in this figure, all arrows correspond to components of fluxes or EMFs.}
\end{figure}

For zone-centred, volume-conservative quantities this procedure follows \citetalias{bc89}.  Thus, for transport in the 1-direction, we have for the flux-corrected quantity, $\widetilde Q$:
\begin{equation}
  \label{eq:case0}
  \begin{split}
  \widetilde{Q}^{N+1}(I,J,K) ~=~ Q^{N+1}(I,J,K) - \frac{1}{\Delta V(I,J,K)} \bigg[ F_{1,Q}^{N+\onehalf}(&I,J,K) \\
  - \sum_{\beta,\eta,\tau=0}^{\nu-1} &f_{1,q}^{n+\tau+\onehalf}\left(i, j+\beta, k+\eta\right)\bigg],
  \end{split}
\end{equation}
where the quantities in square brackets are the flux corrections.  For the purpose of illustration, the coarse zone $(I,J,K)$ is taken to be immediately to the right (increasing $I$) of a fine grid.  $F_{1,Q}^{N+1/2}$ is the time-centred 1-flux of $Q$ (with units $Q {\cal V} \Delta A \Delta t$, where ${\cal V}$ is the coarse velocity)\footnote{Strictly speaking, this is not a flux because of the factor $\Delta t$.  However, for the purposes of accounting, we find it advantageous to define the fluxes with the time steps embedded.} passing across the 1-face co-spatial with the skin of the fine grid, while $f_{1,q}^{n+\tau+1/2}$ are the corresponding fine fluxes (with units $qv\delta A \delta t$, where $v$ is the fine velocity; Figure \ref{fig:reflux}a).  Note the sum over $\tau$ reflecting the fact there are several fine time steps (typically $\nu$ of them) within a single coarse time step.

When flux correcting the face-centred, volume-conserved momenta, we must depart from \citetalias{bc89} as the staggered mesh gives rise to four distinct cases that must be dealt with individually.  The first case is when both the flux and momentum component are parallel to the fine grid skin normal.  Here, the ROI at the boundary is halfway inside the fine grid.  For example, consider the 1-flux of $S_1$ in the ROI straddling the right boundary of a fine grid as shown in Figure \ref{fig:reflux}b where the 1-flux correction takes the form:
\begin{equation}
   \label{eq:case1s1}
   \begin{split}
   \widetilde{S}_1^{N+1}(I,&J,K) ~=~ S_1^{N+1}(I,J,K) - \frac{1} {\Delta V(I,J,K)} \bigg[ F_{1,S_1}^{N+\onehalf}(I-1,J,K) \\
& - \frac{1}{2} \sum_{\beta,\eta,\tau=0}^{\nu-1} \Big(f_{1,s_1}^{n+\tau+\onehalf}\big(i-\tfrac{\nu}{2}-1, j+\beta, k+\eta\big)+f_{1,s_1}^{n+\tau+\onehalf}\big(i-\tfrac{\nu}{2}, j+\beta, k+\eta\big) \Big) \bigg].
    \end{split}
\end{equation}
Note that in the 1-direction, the coarse and fine momenta pass through the same face, but the fluxes do not.  Thus, an average of the fine fluxes at $i-\nu/2-1$ and $i-\nu/2$ is needed to properly centre the fine fluxes, and to ensure local conservation of momentum.

The second case is when the flux is parallel and the momentum component perpendicular to the fine grid skin normal.  Here, the adjacent ROI lies entirely outside the fine grid just as a zone-centred quantity.  For example, consider the 1-flux of $S_2$ in the ROI adjacent the right boundary of a fine grid as shown in Figure \ref{fig:reflux}c, where the flux correction takes the form:
\begin{equation}
   \label{eq:case2s2}
   \begin{split}
   \widetilde{S}_2^{N+1}(I,J,K) ~=~ S_2^{N+1}(I,J,K) - \frac{1}{\Delta V(I,J,K)} &\bigg[ F_{1,S_2}^{N+\onehalf}(I,J,K) \\
   -\!\!\sum_{\beta=-\nu/2}^{\nu/2}\, \sum_{\eta,\tau=0}^{\nu-1}&\, {\cal G}(\beta)\, f_{1,s_2}^{n+\tau+\onehalf}(i, j+\beta, k+\eta) \bigg].
   \end{split}
\end{equation}
The factor $\cal{G}(\beta)$ (equation \ref{eq:gfactor}) ensures that only half of the fine 1-fluxes of the fine ROIs at $\beta=\pm \nu/2$ are included, as a quick glance at Figure \ref{fig:reflux}c will verify.

The third case is when the flux is perpendicular and the momentum component parallel to the fine grid skin normal.  As with the first case, the ROI at the boundary is halfway inside the fine grid.  For example, consider the 1-flux of $S_2$ in the ROI straddling the lower boundary of a fine grid as shown in Figure \ref{fig:reflux}d, where the flux correction takes the form:
\begin{align}
\notag
   \widetilde{S}_2^{N+1}(I,J,K) ~&=~ S_2^{N+1}(I,J,K)\\
   \label{eq:case3s2}
&- \frac{1}{\Delta V(I,J,K)} \Bigg[ \frac{\nu-1}{2\nu}\bigg( F_{1,S_2}^{N+\onehalf}(I,J,K)-F_{1, S_2}^{N+\onehalf}(I+1,J,K)\bigg) \\
\notag
&\quad - \sum_{\beta=1}^{\nu/2}\, \sum_{\eta,\tau=0}^{\nu-1}\, {\cal G} (\beta) \bigg( f_{1,s_2}^{n+\tau+\onehalf}(i,j+\beta,k+\eta)- f_{1,s_2}^{n+\tau+\onehalf}(i+\nu,j+\beta,k+\eta) \bigg)\Bigg].
\end{align}
The fraction $(\nu - 1)/2\nu$ is the \emph{area filling ratio} between the fine to coarse fluxes.  For example, for $\nu=4$, $(\nu - 1)/2\nu = 3/8$ (Figure \ref{fig:reflux}d), and $3/8$ of the coarse flux must be replaced with the overlapping fine fluxes.  Note that $f_{1,s_2} (i,j,k)$ does not contribute to the fine fluxes that replace part of the coarse flux since it is determined, in part, by values from the boundary region of the fine grid, and thus not taken as reliable enough to replace part of the coarse flux determined exclusively from active zones of the coarse grid.

Equation (\ref{eq:case3s2}) assumes that both the left and right faces of the coarse ROI need to be flux corrected.  Should the fine grid in Figure \ref{fig:reflux}d not cover coarse zones $(I,J,K)$ and $(I+1,J,K)$, flux corrections need only be applied to the left side of the coarse ROI $(I,J,K)$.  In this case, the terms $F_{1, S_2}^{N+1/2}(I+1,J,K)$ and $f_{1,s_2}^{n+\tau+1/2}(i+\nu,j+\beta,k+\eta)$ are omitted.  Similarly, if the fine grid does not cover coarse zones $(I-1,J,K)$ and $(I,J,K)$, the terms $F_{1, S_2}^{N+1/2}(I,J,K)$ and $f_{1,s_2}^{n+\tau+1/2}(i,j+\beta,k+\eta)$ are omitted.

The fourth and final case is when both the flux and momentum component are perpendicular to the fine grid skin normal.  Here, no flux corrections are required, as can be readily seen by inspection of Figure \ref{fig:reflux}c.

For the magnetic field components, the operator in the evolution equation (equation \ref{eq:induction}) is the curl, and not the divergence of the hydrodynamical variables.  Thus, $\vec B$ is a surface conserved quantity (rather than volume-conserved) and, as such, we make adjustments to the so-called EMFs (defined below) instead of the fluxes; otherwise the procedure is the same.

Consider the coarse ROI of $B_3$ immediately to the right of a fine grid (Figure \ref{fig:reflux}e).  Following \citet{b01}, we have:
\begin{equation}
  \label{eq:case4}
  \begin{split}
  \widetilde{B}_3^{N+1}(I,J,K) ~=~ B_3^{N+1}(I,J,K) + \frac{1}{\Delta A_3(I,J,K)} &\bigg[ \cale_2^{N+\onehalf}(I,J,K) \\
  &- \sum_{\beta,\tau=0}^{\nu-1} \veps_2^{n+\tau+\onehalf} (i, j+\beta, k) \bigg],
  \end{split}
\end{equation}
where $\cale_2^{N+1/2}(I,J,K) = E_2 \Delta x_2 \Delta t$ is the time-centred coarse 2-EMF\footnote{\label{fn:emfs}Similar to the hydrodynamical fluxes, we define the EMFs with the factor $\Delta t$ embedded to simplify the accounting.} (``electro-motive force") located along the 2-edge, and where $\veps_2^{n+\tau+1/2}(i,j+\beta,k)$ is the time-centred fine 2-EMF, located along the same 2-edge as the coarse 2-EMF.  Here, the quantity in square brackets are the ``EMF corrections", analogous to the flux corrections for the hydrodynamical variables.  In \zeusddd, the EMFs can be evaluated using various algorithms.  In our case, we use the Consistent Method of Characteristics (CMoC) described by \citet{c96}.

Because the magnetic field is an area-conserved quantity, there is no situation where the coarse ROI of a magnetic field component is partially covered by fine ROIs as can happen for the volume-conserved, face-centred momentum.  Thus, equation (\ref{eq:case4}) and its permutations are sufficient to cover all EMF corrections for all field components in all directions.  Note, for example, there are no corrections to be made in the 1-direction for the 1-field, again because of the surface conservative nature of the magnetic field.

Further, induction of magnetic field components penetrating the skin of a fine grid is affected by EMFs computed from quantities taken entirely from within the active portion of the grid.  This is in contrast to a momentum component penetrating the skin, half of whose fluxes are computed from boundary values.  Thus, we take the skin values of the EMFs and the magnetic components they induce to be just as reliable as those evaluated from within the grid, and it is appropriate to restrict the coarse magnetic field values on the skin of a fine grid with the overlying fine values (\eg, equation \ref{eq:over_fcac}).

Finally, flux correction equations for coarse zones adjacent to other sides of a fine grid can be obtained by a suitable permutation of indices and subscripts.

\section{Prolongation}
\label{sec:prolong}
Prolongation is the process by which fine grid zones are filled using the best information available.  This could either be by interpolating values from the underlying coarse grid in a way to ensure local conservation and local monotonicity, or taking them from adjacent or overlapping fine grids. As with restriction, prolongation can be divided into two types.  First, when a fine grid is created or extended, the new fine zones must be filled.  Second, at the beginning of a fine grid time step, fine boundary zones must be set.  Both types require interpolation methods, some of which have been introduced here specifically for static grid refinement.
\subsection{Spatial interpolation}
\label{sub:spatialint}
In an effort to minimise the errors caused by waves travelling across boundaries between fine and coarse grids, we have introduced higher-order interpolation schemes (\eg, piecewise parabolic interpolation, PPI, \citealt{cw84}) to the prolongation step.  This improves the results for adaptive refinement, and has proven important for static refinement where strong waves are required to cross grid boundaries.  By design, these interpolation schemes honour conservation laws and monotonicity.

In all cases and for all variables, we begin by estimating the interface values $Q_{L,R}(I)$ from cubic fits to the coarse grid data (Section 1 of \citealt{cw84}) without the monotonisation or steepening steps.  Now, to fit a parabolic interpolation function, $q_1^*(x)$, across the coarse ROI for zone $I$, we need three constraints.  Two come from requiring that $q_1^*(x)$ passes through $Q_{L,R}(I)$, while the third comes from requiring that the zone-average value $Q(I,J,K)$ times the zone volume equal the volume integral of $q_1^*(x)$.  For the original PPI scheme, this final constraint is written as:
\begin{equation*}
  Q(I,J,K) \Delta x_1(I) ~=\, \int_{\Delta x} q_1^*(x)dx.
\end{equation*}
For our purposes, we need to interpolate fine zone \emph{averages} from the coarse ones and, as such, the conserved quantity is the \emph{Riemann sum} and not the integral.  Thus, in the 1-direction, our final constraint is:
\begin{equation}
  Q(I,J,K) \Delta x_1(I) ~=~ \sum^{\nu-1}_{\alpha=0}\, q_1^*(i+\alpha, j, k) \,\delta x_1(i)
\end{equation}
\begin{equation}
  \label{eq:ppiconserve}
  \Rightarrow \quad Q(I,J,K) ~=~ \frac{1}{\nu} \sum_{\alpha=0}^{\nu-1}\, q_1^*(i+\alpha, j, k)
\end{equation}
for constant $\Delta x$, $\delta x$.  With this, our parabolic interpolation function is (\cf, equation 1.4 in \citealt{cw84}):
\begin{equation}
  \label{eq:1dppi}
    q^*_1(i+\alpha)  ~=~ Q_L(I) + \zeta \big( \, Q_R(I) - Q_L(I) + {\cal H}_1 ( 1 - \zeta ) \big),
\end{equation}
where:
\begin{equation}
   \label{eq:ppifuncs}
\begin{array}{rl}
   \zeta &\!=~ \dfrac{x_1(i+\alpha) - x_1(I)}{\Delta x_1(I)}; \\[12pt]
   {\cal H}_1 &\!=~ \dfrac{1}{f_1(\nu) - f_2(\nu)} \Big( Q(I,J,K) - Q_L(I) - f_1(\nu) \big( Q_R(I) - Q_L(I) \big) \Big);  \\[12pt]
   f_1(\nu) &\displaystyle \!=~ \frac{1}{2\nu^2} \sum_{\xi=1}^{\nu}\, (2\xi - 1) ~=~ \frac{1}{2};\\[15pt]
 f_2(\nu) &\displaystyle \!=~ \frac{1}{4\nu^3} \sum_{\xi=1}^{\nu}\, (2\xi - 1)^2 ~=~ \frac{4\nu^2-1}{12\nu^2}.
\end{array}
\end{equation}
With $q^*_1(i+\alpha)$ determined, we set the differences between the fine and coarse zones:
\begin{equation}
\label{eq:differences}
\delta q_1 (i+\alpha) ~=~ q^*_1(i+\alpha) - Q(I,J,K).
\end{equation}  

For zone-centred quantities, we determine the fine-zone differences in the manner of equation (\ref{eq:differences}) in each of the three directions, and set the interpolated fine zone averages to be:
\begin{equation}
  \label{eq:fineint}
  q(i+\alpha, j+\beta, k+\eta) ~=~ Q(I,J,K) + \delta q_1(i+\alpha) + \delta q_2(j+\beta) + \delta q_3 (k+\eta).
\end{equation}
Given equation (\ref{eq:ppiconserve}) and analogous expressions in the 2- and 3-directions, it is easy to show that this prescription is conservative; that is, \begin{equation}
Q(I,J,K) ~=~ \frac{1}{\nu^3}\,\sum_{\alpha,\beta,\eta=0}^{\nu-1}q(i+\alpha,j+\beta,k+\eta),
\end{equation}
for uniform grids ($\Delta V = \nu^3 \delta V$).

For face-centred quantities, we determine the fine-zone differences in the manner of equation (\ref{eq:differences}) in each of the two orthogonal directions (\eg, in the 2- and 3-directions for $S_1$), and set the interpolated fine zone averages along each coarse face to be:
\begin{equation}
  \label{eq:faceint}
  s_1(i, j+\beta, k+\eta) ~=~ S_1(I,J,K) + \delta s_{1,2}(j+\beta) + \delta s_{1,3} (k+\eta).
\end{equation}
Exactly the same procedure is used to interpolate $B_1$ across each coarse 1-face.  As with the zone-centred quantities, it is easy to show that this procedure conserves momentum (magnetic) flux on the face:
\begin{equation}
  \label{eq:faceconserve}
    S_1(I,J,K) = \frac{1}{\nu^2}\,\sum_{\beta,\eta=0}^{\nu-1}s_1(i,j+\beta,k+\eta).
\end{equation}

The next step is to interpolate the face-centred quantities into the interior of the zone, and it is here where the procedure for momentum and magnetic components diverge.

For the volume-conserved momenta, we perform a linear interpolation between the fine momenta on opposing coarse faces.  For example, between $s_1(i,j+\beta,k+\eta)$ and $s_1(i+\nu,j+\beta,k+\eta)$ we set:
\begin{equation}
  \label{eq:momlinint}
  s_1(i+\alpha,j+\beta,k+\eta) ~=~ (1 - \zeta)\, s_1(i,j+\beta,k+\eta) + \zeta\, s_1(i+\nu,j+\beta,k+\eta),
\end{equation}
where $\alpha=1,\dots,\nu-1$ and $\zeta=\alpha\delta x/\Delta x$.  This prescription guarantees that over a \emph{zone-centred volume}, the prolongation of the 1-momentum is conservative.  The fact that conservation is over the zone-centred volume and not specifically the 1-momentum ROI is required for equations (\ref{eq:faceconserve}) and (\ref{eq:momlinint}) to be consistent with the restriction procedure of equation (\ref{eq:case1s1}).

For the area-conserved magnetic field, a simple linear interpolation between opposing coarse faces does not preserve the solenoidal condition.  Thus, we turn to a generalised, directionally unsplit version of the algorithm described by \citet{ll04} in which $\nabla \cdot \vec b=0$ so long as $\nabla\cdot\vec{B} = 0$ holds on the underlying coarse grid.

\begin{figure}[t!]
  \begin{center}
    \includegraphics[width=0.85\textwidth]{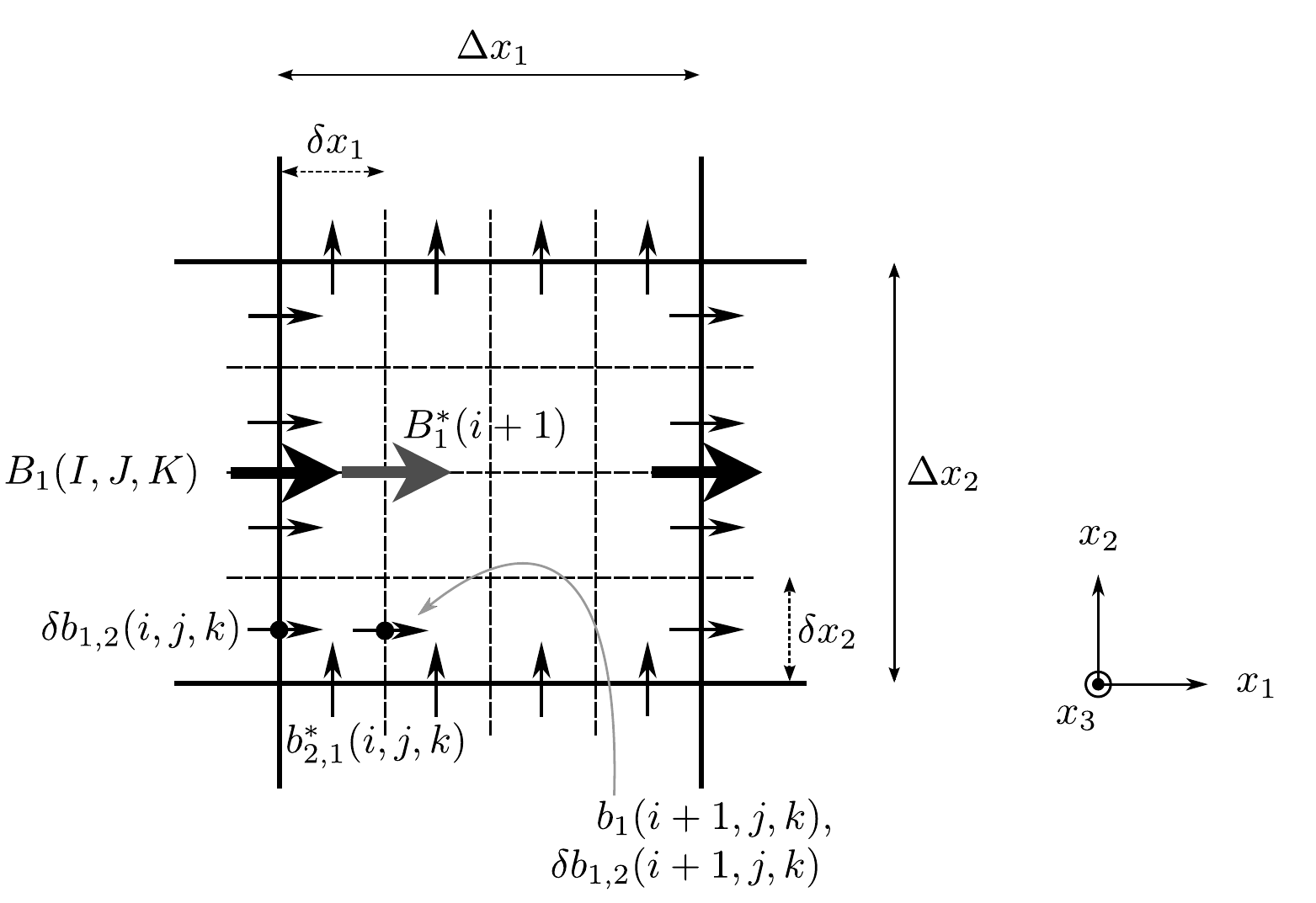}
  \end{center}
  \caption{\label{fig:lili} Schematic representation of the directionally unsplit \citeauthor{ll04} algorithm for calculating fine values of $\vec{b}$ between coarse grid faces when $\nu = 4$.}
\end{figure}

In this approach, we take the coarse and recently interpolated fine values of magnetic field (equation \ref{eq:faceint}), and apply the solenoidal condition to determine ``intermediate'' coarse field values (\eg, $B_1^*$; Figure \ref{fig:lili}), which are cospatial with fine zone faces.  For example, the intermediate values for $B_1$ are given by:
\begin{equation}
  \label{eq:lilistep1}
  \begin{split}
    B_1^*(i+\alpha+1) = B_1^*(i+\alpha) - \delta x_1(i+\alpha) \Bigg(& \frac{b_{2,1}^*(i+\alpha,j+\nu,k) - b_{2,1}^*(i+\alpha,j,k)}{\Delta x_2(J)}\\ + \, & \frac{b_{3,1}^*(i+\alpha,j,k+\nu) - b_{3,1}^*(i+\alpha,j,k)}{\Delta x_3(K)}\Bigg),
  \end{split}
\end{equation}
where:
\begin{align}
  \label{eq:lilifuncs}
  b_{2,1}^*(i+\alpha,j,k) = & \,\, \frac{1}{\nu} \sum_{\eta = 0}^{\nu-1}\, b_2(i+\alpha,j,k+\eta), \nonumber \\
  b_{3,1}^*(i+\alpha,j,k) = & \,\, \frac{1}{\nu} \sum_{\beta = 0}^{\nu-1}\, b_3(i+\alpha,j+\beta,k); \\
  B_1^*(i+0) = & \, B_1(I,J,K), \nonumber
\end{align}
and $0 \leq \alpha \leq \nu-2$.  Given $B_1^*(i+\alpha+1)$, and the differences between fine and coarse values at the coarse faces ($\delta b_{1,l},l=2,3$; equation \ref{eq:differences}), we then calculate the fine magnetic field at $i+\alpha+1$:
\begin{equation}
  \label{eq:lilistep2}
  \begin{split}
    b_1(i+\alpha+1,j+\beta,k+\eta) = B_1^*(i+\alpha+1)\, + \, & \delta b_{1,2}(i+\alpha+1,j+\beta,k) \\ +\, & \delta b_{1,3}(i+\alpha+1,j,k+\eta),
  \end{split}
\end{equation}
where:
\begin{equation}
  \label{eq:lililinear}
  \begin{split}
    \delta b_{1,2}(i+\alpha,j+\beta,k) =~ & (1 - \zeta)\, \delta b_{1,2}(i,j+\beta,k) + \zeta\, \delta b_{1,2}(i+\nu,j+\beta,k), \\
    \delta b_{1,3}(i+\alpha,j,k+\eta) =~ & (1 - \zeta)\, \delta b_{1,3}(i,j,k+\eta) + \zeta\, \delta b_{1,3}(i+\nu,j,k+\eta),
  \end{split}
\end{equation}
and $\zeta=\alpha\delta x/\Delta x$.

Proceeding incrementally from $i+1$ to $i+\nu-1$ provides all of the values of $b_1(i,j,k)$ between the coarse faces $(I,J,K)$ and $(I+1,J,K)$.  Values for the other magnetic field components are given by a straightforward permutation of indices.  In aggregate, these yield a third-order interpolation of fine magnetic field components that preserve the divergence of the underlying coarse magnetic field to machine round-off error.

Finally, a note on the use of vector potentials.  \citet{in02} have demonstrated that the vector potential, $\vec A$, may be used in the CT algorithm (\citealt{eh88}) instead of the magnetic field with results identical to machine round-off error.  Thus, one may be tempted to adopt this approach so that prolongation methods simpler than the \citeauthor{ll04} algorithm may be used on $\vec A$ as a way to guarantee that the fine grid satisfies the solenoidal condition.  Reasons for \emph{not} taking this approach are given in Appendix \ref{app:vecpot}.
\subsection{Temporal interpolation}
\label{sub:timeint}
Fine grids require prolonged boundary values at the beginning of each fine time step, and thus we also need to perform temporal interpolations on the coarse values.  Following \citetalias{bc89}, we perform a linear interpolation on the coarse hydrodynamic variables in time, and then spatially interpolate them as described above to obtain the necessary boundary information for each fine time step.

For the magnetic field, the fine components penetrating the skins of the fine grid are retained; only those values completely within the boundary region are prolonged from the coarse grid.  For this, we require coarse grid EMFs on the fine grid skin and within the boundary region that are time centred on the current fine time step.

The coarse EMFs coincident with the fine grid skin are replaced with the spatial \emph{and} temporal sum of the overlying fine EMFs, where the sum over time is necessary since the time step is embedded in our definition (footnote \ref{fn:emfs}).  Thus and for example, for $\cale_2$ we set:
\begin{equation}
\label{eq:interemf2}
  \cale_2^{N+\psi}(I,J,K) ~=~ \sum_{\tau=0}^{\tau '}\,\sum_{\beta=0}^{\nu-1}\, \veps_2^{n+\tau+\onehalf} (i,j+\beta,k),
\end{equation}
where $\psi=(2\tau'+1)/2\nu$, and where $0 \leq \tau ' \leq \nu-2$ designates the last completed fine time step.  This yields values on the fine grid skin which are time-centred between the beginnings of the coarse time step and the current fine time step, and take into account the superior quality of the fine grid data.

The coarse EMFs residing entirely within the fine grid boundary region are retained.  However, these are time-centred for the coarse grid and, as such, have embedded in them a factor $\Delta t$.  To render them coeval with fine time step $\tau^\prime$ and thus the coarse skin EMFs computed in equation (\ref{eq:interemf2}), we must multiply them by $(2\tau^\prime+1)/\nu=2\psi$ to correct the embedded time step factor.

With coarse EMFs on the fine skin and inside the fine boundary region properly time-centred, the coarse magnetic field components are updated to the beginning of the current fine time step using equation (\ref{eq:induction}), and it is these values that are used in the prolongation methods described in Section \ref{sub:spatialint} to create the fine grid magnetic field boundary values.  

One might ask why a linear temporal interpolation of the divergence-free coarse magnetic field components isn't enough to preserve the solenoidal condition within the boundary region.  Indeed, such an approach does guarantee $\nabla\cdot\vec B=0$ \emph{within the region interpolated}, but not in the layer between the interpolated region and its immediate non-interpolated neighbours.  The method outlined above based on the coarse EMFs allows interpolations to be performed locally and only where they are needed, while still preserving the solenoidal condition globally to machine round-off error.

\subsection{Monotonicity}
\label{sub:mono}
We have found it necessary to maintain a certain monotonicity in our prolongations.  Failure to do so can lead to negative pressures (even when solving the internal energy equation) and violations of the CFL condition.  For example, if the current time step is governed by the sound speed, and the prolongation process leads to an interpolated density less than the surrounding zones, the sound speed in a fine zone could be greater than that which was used in determining the CFL time step, possibly leading to numerical oscillations and loss of stability.

Sequentially stringing together two (or three) 1-D PPIs as we do for prolongation of zone-centred quantities (\eg, equation \ref{eq:fineint}) can lead to non-monotonic behaviour even if each 1-D interpolation is separately monotonic.  If a PPI-determined value $q(i+\alpha,j+\beta,k+\eta)$ is found to lie outside the range set by neighbouring coarse zones:
\begin{equation*}
\left. \begin{array}{lcll}
  Q_{\min} \!&=&\! {\rm min} &\!\! \big( Q(I+\Gamma,J+\Lambda,K+\Upsilon) \big); \\[6pt]
  Q_{\max} \!&=&\! {\rm max} &\!\! \big( Q(I+\Gamma,J+\Lambda,K+\Upsilon) \big), \end{array}\right\} ~ -1 \leq \Gamma, \Lambda, \Upsilon \leq +1,
\end{equation*}
then we ``fall back'' to piece-wise linear interpolations (PLI; \citealt{vL77}, Appendix \ref{app:curvi}).  In rare cases where the PLI-determined value is also non-monotonic (possible only in 3-D), we revert to piece-wise constant interpolations (PCI, \emph{a.k.a.}\ ``direct injection" or ``donor cell"):
\begin{equation*}
  \label{eq:limiter}
  q(i+\alpha,j+\beta,k+\eta) ~=~ Q(I,J,K).
\end{equation*}
Where PPI yields non-monotonic results, we note that transmission of strong waves across changes in resolution (\eg, static grids) is significantly improved if one first tries PLI rather than falling back directly to PCI or limiting the interpolated value to lie between $Q_{\min}$ and $Q_{\max}$.

\section{Boundary conditions}
\label{sec:bcs}

Boundary conditions are applied directly to the hydrodynamical variables ($\rho$, $e$, $\vec v$) and indirectly to the magnetic field via the EMFs.  Attempting to apply boundary conditions directly to $\vec B$ often generates monopoles in the boundary regions, which can have significant effects on the dynamics in the active grid, as illustrated in Appendix \ref{app:monopole}.
 
In addition to the usual boundary conditions applied by the \zeus module during each MHD cycle, the AMR module must also set boundary conditions on two occasions.  After restriction and before prolongation, boundary values are set on the coarse grid.  Then, after prolongation and before the \zeus module is called for the next MHD cycle, boundary values for the fine grid are set using the results of the prolongation step. These additional applications of boundary conditions are necessary to maintain the imposed physical boundaries after restriction and prolongation have altered some of the values on the active grid, and to reconcile boundaries of grids that may be contained within or adjacent to other grids (henceforth referred to as \emph{adjacent boundaries}).

Because of their nature, physical boundaries (inflow, outflow, reflecting; those traditionally applied by single-grid MHD codes), and adjacent boundaries must be treated differently, and each is discussed in turn.

\subsection{Physical Boundaries}
\label{sub:physbcs}

Generally, a fine grid is completely embedded within a coarse grid. The single exception is when both grids share a physical boundary and two items of note must be borne in mind when adapting the physical boundary condition routines in \zeusddd to \azeus.

First, since each grid has only two boundary zones, only part of the coarse boundary region is covered by fine boundary zones.  Thus, coarse boundary zones cannot be included in the restriction step; a particular concern when setting magnetic boundary conditions.  In \azeus, we extend the EMF correction scheme of Section \ref{sub:reflux} by retaining \emph{three} layers of transverse EMF corrections, and \emph{two} layers of longitudinal EMF corrections\footnote{\eg, the transverse (longitudinal) EMFs for the 1-boundary are $\varepsilon_2$ and $\varepsilon_3$ ($\varepsilon_1$).}, including the skin layer plus two additional layers interior and adjacent to the skin.  Immediately before the restriction step, physical boundary conditions are applied to the EMF corrections, which are then used to update the boundary values of the magnetic field components in the coarse grid according to equation (\ref{eq:case4}). Done in this fashion, there is no risk of introducing monopoles to the coarse boundary zones during the restriction step when two or more grids overlap a physical boundary.

Second, not handled properly, inflow boundary conditions can introduce unexpected violations of conservation laws which can cause unwanted discontinuities in the boundary.  In particular, if a boundary variable is to be set according to an analytical function of the coordinates, that variable should be set to the \emph{zone average} of that function, and not simply to the function value at the location of the variable.  While this is good advice for a single grid application, it is critical for AMR.

\begin{figure}[t!]
  \begin{center}
    \includegraphics[width=0.35\textwidth]{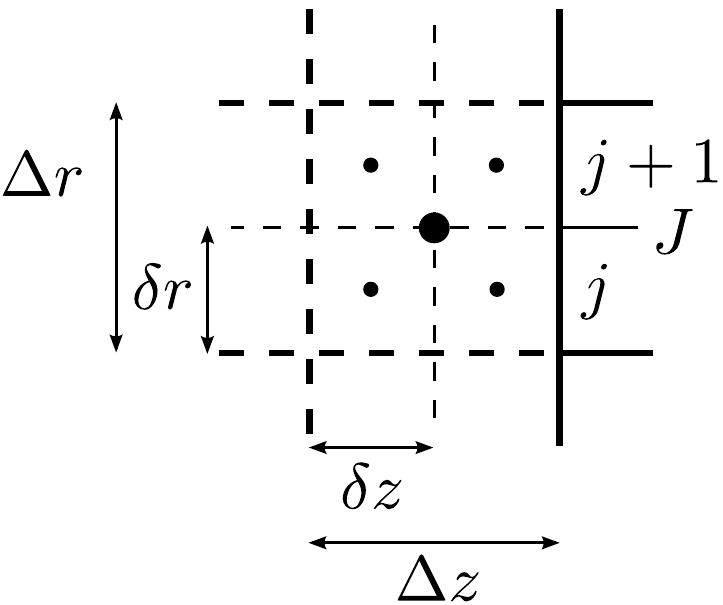}
  \end{center}
  \caption{\label{fig:inflowBC} A single coarse zone in the $z=0$ boundary with zone centre at $r(J)=r$ and with refinement ratio of $\nu = 2$.}
\end{figure}

For example, suppose the density profile:
\begin{equation}
\label{eq:hystat}
\rho(r) ~=~ \frac{1}{r^{3/2}},
\end{equation}
is to be maintained in cylindrical coordinates along the $z=0$ boundary.  Let $\rho(J)$ be the density in the coarse zone of dimension $(\Delta z,\Delta r)$ centred at $r(J)=r$, and let $\rho(j)$ and $\rho(j+1)$ be the densities in the fine zones of dimension $(\delta z,\delta r)$ centred at $r(j)$ and $r(j+1)$ (Figure \ref{fig:inflowBC}).  For a refinement ratio of 2, $\Delta r = 2\delta r$, $\Delta z = 2\delta z$, $r(j)=r-\onehalf\delta r$, and $r(j+1)=r+\onehalf\delta r$.

If we na\"{\i}vely set $\rho(J)$ using equation (\ref{eq:hystat}), the mass of zone $J$ (with volume $\Delta V=r\Delta r \Delta z$) is:
\begin{equation*}
  m(J) ~=~ \rho(J) \Delta V(J) ~=~ \frac{\Delta r \Delta z}{r^{1/2}}.
\end{equation*}
Similarly, for the fine zones,
\begin{equation*}
  m(j) ~=~ \frac{\delta r \delta z}{\big(r-\onehalf \delta r\big)^{1/2}}; \qquad m(j+1) ~=~ \frac{\delta r \delta z}{\big(r+\onehalf\delta r\big)^{1/2}}.
\end{equation*}
Adding the masses in the four fine zones contained by the coarse zone, we find:
\begin{equation*}
  2\big(m(j)+m(j+1)\big) ~=~ m(J)\!\left( 1 + \frac{3}{32} \!\left(\frac{\delta r}{r}\right)^{\!\!2} + \dots \right) \,>~ m(J),
\end{equation*}
and the physical boundary conditions violate mass conservation.

Instead, $\rho(J)$ should be set to the \emph{zone average}, determined by:
$$\rho(J) ~=~ \frac{m(J)}{\Delta V(J)}$$
where the mass function, $m(J)$, is given by:
\begin{equation}
\label{eq:mofj}
m(J) ~=\, \int_{\Delta V(J)}\!\!\rho(r) dV,
\end{equation}
where $\rho(r)$ is the given density profile (\eg, equation \ref{eq:hystat}).  Similar expressions apply for $m(j)$, and $m(j+1)$.  Done in this fashion, it is easy to show that
$$\rho(J)\Delta V(J) ~=~ 2\big( \rho(j+1)\delta V(j+1) + \rho(j)\delta V(j) \big),$$
and both grids agree on the mass contained within coarse boundary zone $J$ to within machine round-off error.  With a little bit of algebra, this can be confirmed analytically for the specific case of equation (\ref{eq:hystat}).  For more complicated profiles, a numerical integrator can be employed to perform the necessary integrations in equation (\ref{eq:mofj}).

\subsection{Adjacent Boundaries}
\label{sub:selfcompute}
Unlike Godunov methods which typically require a single application of boundary conditions at the end of each MHD cycle, the operator-split nature of \zeus requires boundary conditions to be set several times.  For physical boundaries, this is not an issue; physical boundary conditions may be set whenever needed based on data from a single grid.  However, adjacent boundaries pose a unique challenge in \azeus since the \zeus module is aware only of the grid being updated, and other grids cannot be accessed in the middle of an MHD cycle to update these boundaries. 

We have introduced into \azeus the concept of ``self-computing" boundary conditions for adjacent boundaries.  In this approach, boundary zones are set at the beginning of each time step using the best information available (either from an adjacent grid of the same resolution, or from the prolongation of an underlying coarse grid possibly interpolated in time), and then the full set of operator-split MHD equations are applied to both the boundary and active zones; no adjacent boundary zones are reset to any pre-determined quantity inside a single MHD time step.  Further, self-computed boundary zones are included in the calculation of the CFL time step, which we find critical (for an operator-split code such as \azeus) in minimising transmission errors when waves of any significant amplitude cross adjacent boundaries.

Of course, assumptions about missing data beyond the outermost edge of the grid must be made, and ``pollution'' from these missing data necessarily propagates inward.  For example, consider the left 1-boundary where $v_1(i = 1)$ and $v_1(i = 2)$ represent the boundary values of $v_1$, $v_1(i = 3)$ the skin value (treated as a boundary value), and $v_1(i = 4)$ the first active value.  The pressure gradient at $v_1(1)$ is proportional to $p(1)-p(0)$, yet $p(0)$ is completely unknown [$p(1)$ and $p(2)$ are the only boundary values available].  The best we can do for a missing datum such as this is to extrapolate assuming a zero-gradient, in which case $p(0) = p(1)$ and no pressure gradient is applied to $v_1(1)$.  In this manner, $v_1(1)$ is polluted by the missing datum at the very beginning of the source step.

Additional steps inside a single MHD cycle include the application of artificial viscosity, adiabatic expansion/com\-pression for the internal energy equation, the transport steps, and the induction step. All but the latter contribute to propagating pollution from missing data toward the active portion of the grid.  Ideally, one would carry enough boundary zones to prevent pollution from reaching the active grid within a single MHD step, so that the AMR module (with knowledge of all grids) can reset all boundary values to new and unpolluted values before their effects ever reach the active zones.  However, for \citet{vL77} interpolation in the transport step and within a single MHD cycle, pollution from missing data reaches the skin and first active zone-centre on the grid, as well as the first active face and second active zone-centre if the local velocity points away from the boundary (true regardless of which energy equation is used).  Thus, to completely prevent pollution from reaching the active grid, we would have to double the number of boundary zones from two to four.

We have investigated this effect thoroughly and have found no test problem that demonstrates anything but the slightest quantitative effect from pollution by missing data at adjacent boundaries between fine and coarse grids.  For adjacent boundaries between grids of like resolution, we have elected to force these grids to overlap (for reasons explained in Section \ref{sec:gcreate}) by an amount sufficient to eliminate the problem of missing data pollution altogether.

\section{Grid creation and adaptation}
\label{sec:gcreate}
Creation or modification of adaptive grids in \azeus proceeds in a manner similar to \citet{bbsw94} including the suggestions of \citet{br91}, with two important differences.  First, we have had to modify the proper nesting criterion of \citetalias{bc89} by increasing from one to three the number of zones at level $l-1$ separating an active zone at level $l$ from level $l-2$.  This is because prolongation of boundary values for level $l$ from level $l-1$ requires one zone from level $l-1$ beyond the edge of a grid at level $l$, plus two additional zones at level $l-1$ on either side to satisfy the five-zone molecule needed by PPI.

\begin{figure}[t!]
  \begin{center}
    \includegraphics[width=1.0\textwidth]{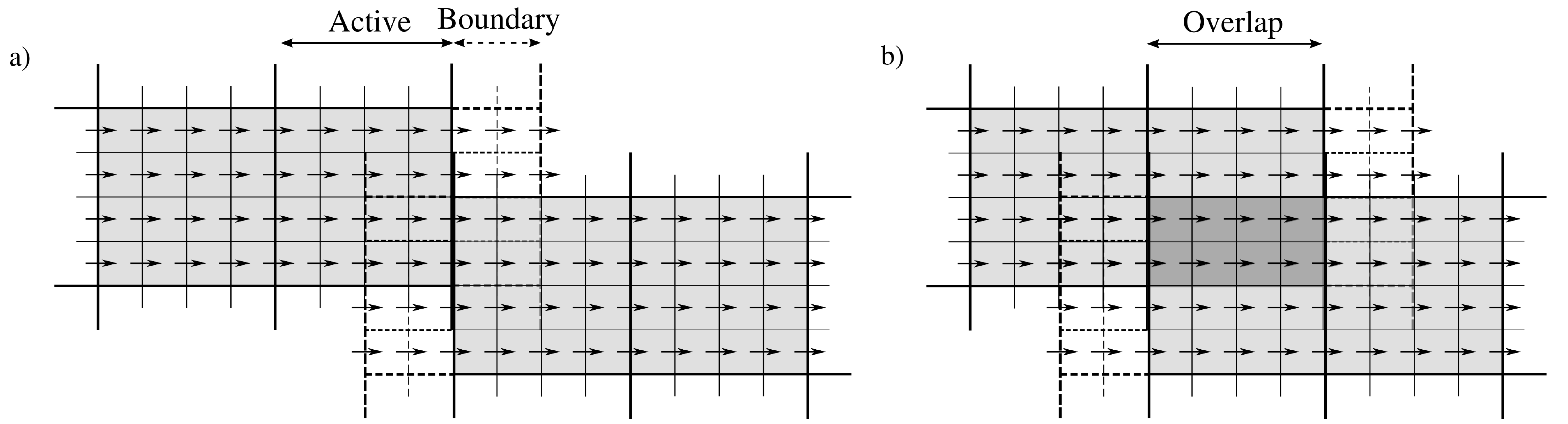}
  \end{center}
  \caption{\label{fig:overlap} Two grids that originally abut (panel a) are made to overlap by at least one coarse zone (panel b).}
\end{figure}

Second, \citetalias{bc89} allow grids of the same resolution to abut without overlapping, whereas we have found it advantageous for at least two reasons to extend abutting grids so that they overlap by a minimum of one coarse zone (Figure \ref{fig:overlap}). For one, since the momenta penetrating a grid skin are no more reliable than boundary values, two abutting grids would, in general, disagree on the values of the momentum penetrating their common skin.  This turns out to pose an intolerable ambiguity in the solution.  Second, the problem of pollution propagating from missing data onto the active grid (Section \ref{sub:selfcompute}) is completely averted by overlapping two abutting grids by one coarse zone.

By forcing abutting grids to overlap, there is always a clear choice of which value to use at a given location.  Where the boundary and skin values of one grid overlap the active zone values of another, the active zone values prevail, and are used by the other grid for its boundary values.

Grids which are abutting at the end of the grid generation and modification process, but before they are prolonged, are made to overlap by one coarse zone.  Numerous situations arise in which more than one grid may overlap at the same location (\eg, the test problems in sections \ref{ssub:blast} and \ref{ssub:otvortex}), and this can result in overlaps which are greater than one coarse zone.  Further, the possibility of a complex distribution of AMR grids opens up a whole host of pedagogical cases which one must consider, particularly when looking to maintain the solenoidal condition.  During the development of \azeus, we have carefully examined each case involving multiple levels with numerous grids overlapping each other, to ensure that where grids overlap, they all agree on the flow variables and all conservation laws are preserved to machine round-off error.

In this regard, the EMFs turn out to be a very sensitive discriminator.  Any mismatches between overlapping grids in the conserved variables at the beginning of a time step will first generate mismatched velocities, followed by disagreements in co-spatial EMFs before $\nu$ time steps have passed.  These can be detected, for example, as monopoles arising in either the boundary or active zones in either or both of the overlapping grids.  Even if such monopoles are restricted entirely to the boundary zones, their unphysical forces can affect neighbouring active zones whose effects propagate rapidly throughout the grid (\eg, Appendix \ref{app:monopole}). Preserving agreement between overlapping zones including boundary zones to machine round-off is, therefore, of paramount importance.

\section{Numerical Tests}
\label{sec:tests}
We have verified \azeus against a number of standard test problems, some of which are presented in this section.  Further results will be posted to \href{http://www.ica.smu.ca/azeus}{\texttt{http://www.ica.smu.ca/azeus}}, as they become available.  Additional and similar test problems for \zeusddd (without AMR) are found in \citet{c10} and on-line at \href{http://www.ica.smu.ca/zeus3d}{\texttt{http://www.ica.smu.ca/zeus3d}}.
\subsection{1-D shock tubes}
\label{sub:1d}
While most AMR applications rely exclusively on dynamic grids, certain applications, particularly those which exhibit some degree of self-similarity, can benefit enormously from the use of a base of nested, static grids (\eg, \citealt{rc11}).  As such solutions evolve, waves of all types---including strong shocks---must pass sequentially from one grid to another, and it is here where our higher order prolongation algorithms for adjacent boundaries are critical.  

\azeus has been tested with all 1-D shock tube problems from \citet[\citetalias{rj95}]{rj95} for both static and dynamic grids, and we show the results from two to highlight these features of the code.  Both tests use the total energy equation with $\gamma = 5/3$, ${\cal C} = 0.75$ (Courant number), and artificial viscosity parameters \verb|qcon| = 1.0 and \verb|qlin| = 0.2.

\begin{figure}[t!]
  \begin{center}
    \includegraphics[width=0.75\textwidth]{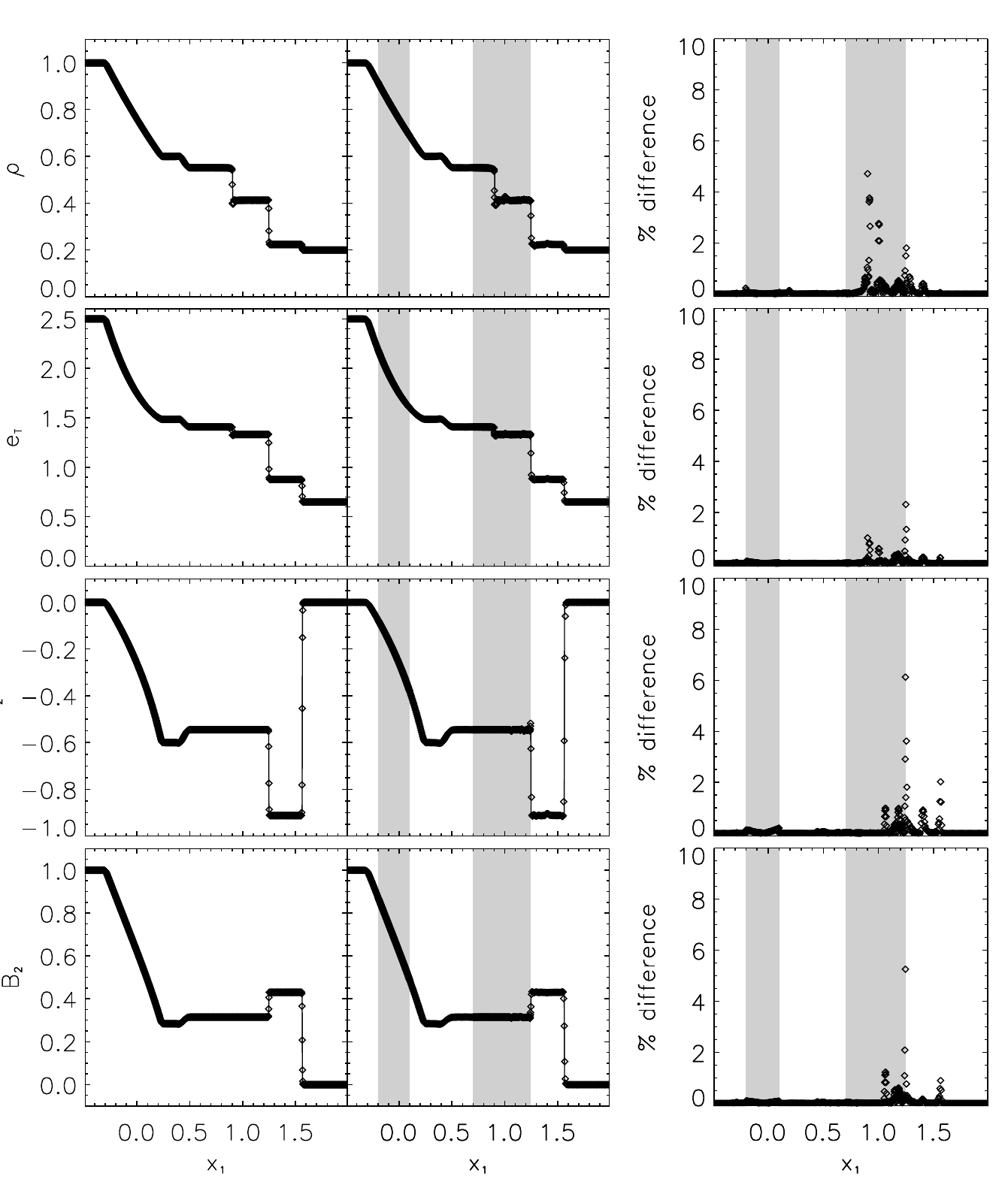}
  \end{center}
  \caption{\label{fig:testrj4a} Static grid solution to problem (4a) of \citetalias{rj95} at time $t = 0.45$.  The initial left and right states are $(\rho, v_1, v_2, v_3, B_2, B_3, p) = ( 1, 0, 0, 0, 1, 0, 1)$ and $(0.2, 0, 0, 0, 0, 0, 0.1)$, with $B_1 = 1$.  From left to right, the physical features are: (1) fast rarefaction, (2) slow rarefaction, (3) contact discontinuity, (4) slow shock, and (5) ``switch-on'' shock.  \textit{Left panels:} uniform ``fine'' grid solution; \textit{middle panels:} same solution with two fine, static grids (gray) overlying the coarse grid; \textit{right panel:} percent difference between the uniform and static grid solutions.  Solid lines are the analytical solutions from the Riemann solver described in \citetalias{rj95}.}
\end{figure}

For static grids, Figure \ref{fig:testrj4a} shows two solutions for $\rho$, $e_{\rm T}$, $v_2$, and $B_2$ from problem (4a) of \citetalias{rj95} over a domain of $x_1 \in [-0.5,2.5]$ (three times larger than \citetalias{rj95}) for a time $t = 0.45$ (three times longer).  The initial discontinuity is placed at $x_1 = 0.5$, with the left and right states given in the figure caption.  The left panels show the domain resolved with 1200 zones and no AMR, while the central panels show the AMR solution with a base resolution of 600 zones and fine static grids with a refinement ratio of 2 placed at $x_1 \in [-0.2, 0.1]$ and $x_1 \in [0.7,1.245]$ (grey).  These locations allow all the physical features, with the exception of the sluggish slow rarefaction, to suffer at least one change in resolution. The right panels show the percent differences between the two solutions.

Discounting all zones trapped within a discontinuity (which, even without AMR, are are already in error by as much as 100\% since discontinuities are supposed to be infinitely sharp), the maximum error one can attribute to the use of static grids in any of the variables is less than 1\%.  As further evidence of the ability of \azeus to pass waves of all types across adjacent boundaries, \citet{rc11} find almost no sign of reflection, refraction, or distortion of any type of wave across any of the static grid boundaries in their 2-D axisymmetric simulations of protostellar jets in cylindrical coordinates.

\begin{figure}[t!]
  \begin{center}
    \includegraphics[width=0.9\textwidth]{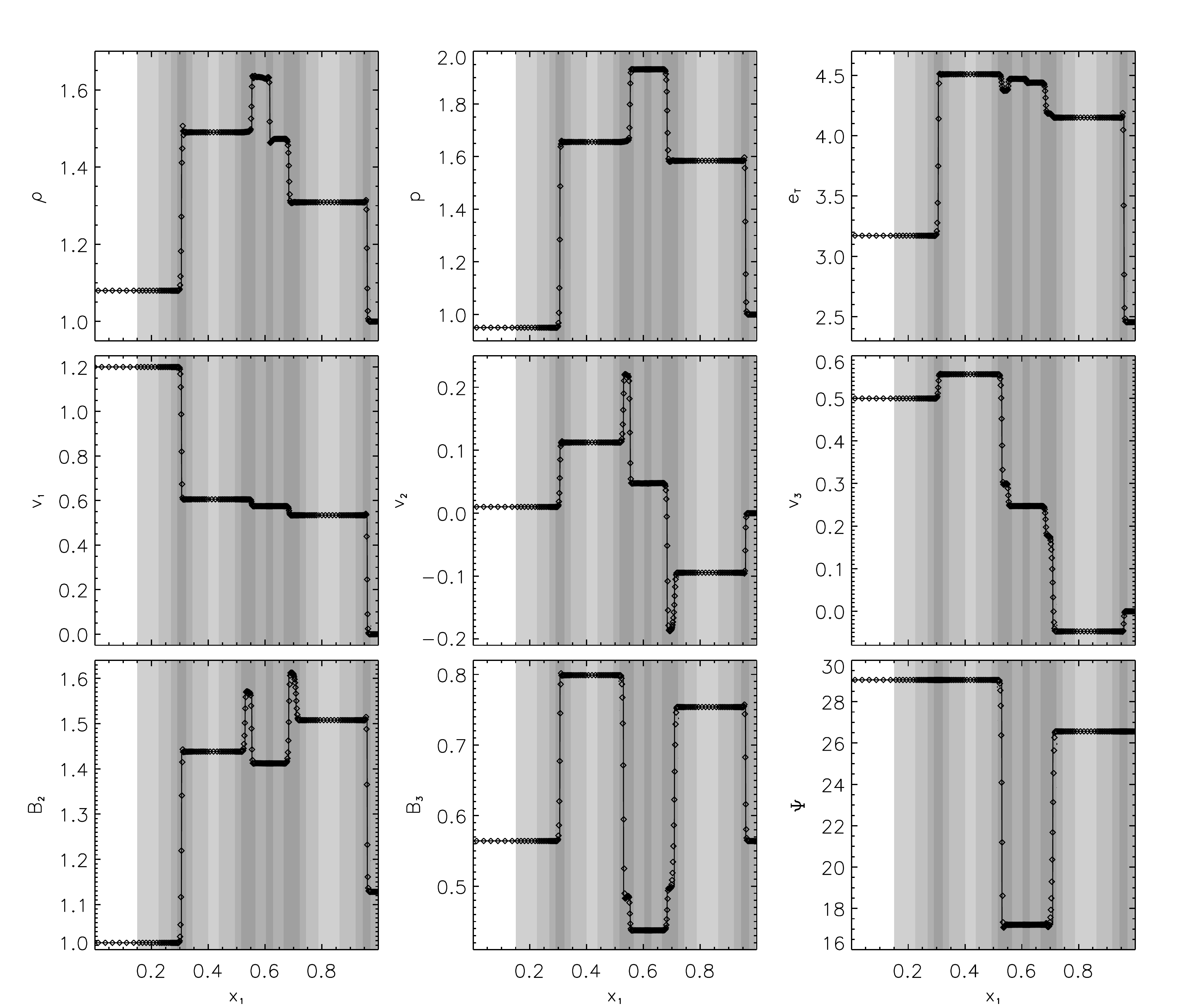}
  \end{center}
  \caption{\label{fig:testrj2a} \azeus solution to problem (2a) of \citetalias{rj95} at $t = 0.20$.  The initial left and right states are $(\rho, v_1, v_2, v_3, B_2, B_3, p) = (1.08, 1.2, 0.01, 0.5, 3.6/\sqrt{4\pi}, 2/\sqrt{4\pi}, 0.95)$ and $(1, 0, 0, 0, 4/\sqrt{4\pi}, 2/\sqrt{4\pi}, 1)$, with $B_1 = 2/\sqrt{4\pi}$.  The physical features, from left to right, are: (1) fast shock, (2) rotational discontinuity, (3) slow shock, (4) contact discontinuity, (5) slow shock, (6) rotational discontinuity, and (7) fast shock.  Shaded regions indicate the location of finer grids, with the level of shading indicating the level of refinement. $\Psi = \tan^{-1} (B_3/B_2)$ is the angle between the transverse field components.  The solid lines are the analytical solution from the Riemann solver described in \citetalias{rj95}.}
\end{figure}

For dynamic grids, Figure \ref{fig:testrj2a} illustrates the results of test problem (2a) of \citetalias{rj95}.  We employ four levels of refinement (sequentially darker shades of grey) above the base grid with a resolution of 40 zones over the domain $x_1 \in [0.0, 1.0]$.  The initial discontinuity is located at $x_1 = 0.5$, with the left and right states given in the figure caption. To track the developing features, refinements are based on three criteria \citep{k98}: contact discontinuities (CD), shocks, and gradients in certain variables, each requiring a threshold value to be set above which a zone is flagged for refinement.  Here, we set parameters \verb|tolcd| = \verb|tolshk| = 0.1 for CDs and shocks, respectively, and check $v_2$ for gradients above \verb|tolgrad| = 0.1.  For this problem, the gradient detector is needed to refine both the discontinuity of the initial conditions as well as rotational discontinuities ($x_1 \simeq 0.53$ and $x_1 \simeq 0.71$ in Figure \ref{fig:testrj2a}), neither of which are detectable as a CD or shock.

Additional parameters used to obtain this solution include \verb|kcheck| = 5 (the number of cycles between successive grid modifications), \verb|geffcy| = 0.95 (the minimum allowed grid efficiency, defined as the ratio of zones flagged for refinement to the number of zones actually present in a new grid), and \verb|ibuff| = 2 (the number of buffer zones added around flagged zones).  These parameters, in addition to the non-AMR parameters, are summarised in Table \ref{tab:codeparam}.

\begin{table}[!t]
  \begin{center}
  \begin{tabular}{cp{0.65\textwidth}}
    \tableline
    parameter & description \\
    \tableline\tableline
    ${\cal C}$ & the Courant number \\
    \verb|qcon| & quadratic artificial viscosity parameter\tablenotemark{a} \\
    \verb|qlin| & linear artificial viscosity parameter\tablenotemark{a} \\
    \verb|tolcd| & threshold value for the CD detector (range: 0 -- 1) \\
    \verb|tolshk| & threshold value for the shock detector (range: 0 -- 1) \\
    \verb|tolgrad| & threshold value(s) for the gradient detector(s) (range: 0 -- 1) \\
    \verb|kcheck| & number of cycles between successive grid modifications \\
    \verb|geffcy| & the minimum allowed fractional grid efficiency for creation of new grids (range: 0 -- 1) \\
    \verb|ibuff| & the number of buffer zones added around zones flagged for refinement \\
    \tableline
  \end{tabular}
  \end{center}
  \tablenotetext{a}{see \citealt{c10} for a more detailed description of the artificial viscosity parameters.}
  \caption{\label{tab:codeparam} Summary and description of the significant user-set parameters in \azeus.}
\end{table}
\subsection{2-D tests}
\label{sub:2d}
\subsubsection{MHD Blast}
\label{ssub:blast}
The MHD blast problem of \citet{ldz00} and \citet{gs05} has proven to be a very valuable test of our AMR algorithms and for rooting out problems in the code.  It is also a good test of directional biases and the ability of a code to handle the evolution of strong MHD waves.  We initialise the problem in the same manner as \citet{c10} with domain $x \in [-0.5,0.5], y \in [-0.5,0.5]$ and $(\rho, \vec{v}, B_1, B_2, B_3) = (1, 0, 5\sqrt{2}, 5\sqrt{2}, 0)$ everywhere.  Within radius $r = 0.125$ of the origin, we set the gas pressure to $p = 100$, and $p = 1$ elsewhere.  

\begin{figure}[t!]
  \begin{center}
    \includegraphics[width=1.0\textwidth]{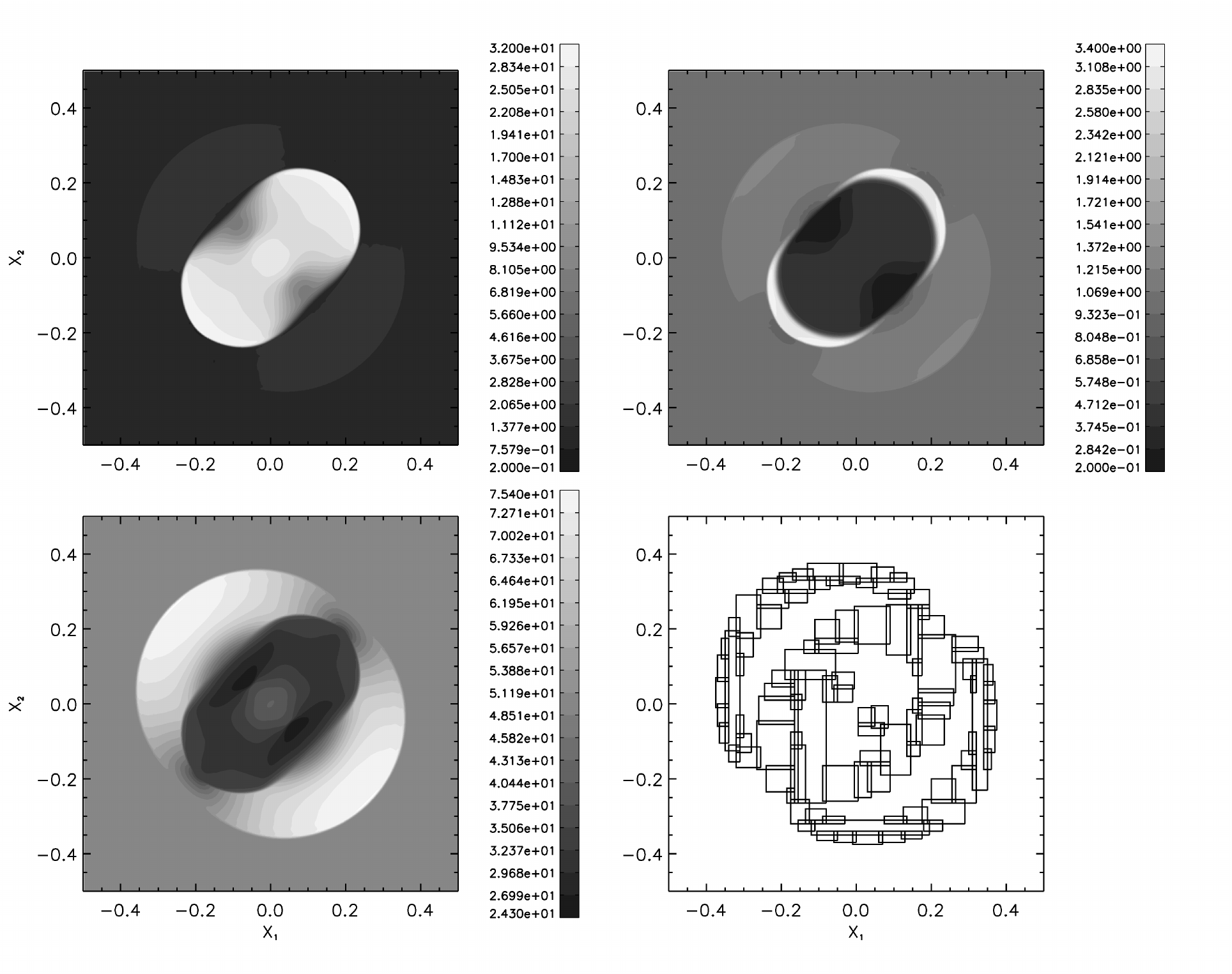}
  \end{center}
  \caption{\label{fig:blast2d} \azeus solution for the 2-D MHD blast problem at $t = 0.02$ using 2 levels of refinement.  \textit{Top left:} gas pressure;  \textit{bottom left:} magnetic pressure ($p_B = |B|^2/2$); \textit{top right:} gas density; and  \textit{bottom right:} distribution of AMR grids at $t = 0.02$.}
\end{figure}

Figure \ref{fig:blast2d} presents our results for the 2-D blast problem at $t = 0.02$ for a base grid of $200^2$ zones, and one additional level of refinement with ratio $\nu = 2$.  For this test, we have purposely chosen a high grid efficiency of \verb|geffcy| = 0.92 to strenuously test the ability of \azeus to handle a large number of grids in a complicated pattern and refine on features which are not preferentially aligned with a coordinate axis.  Typically, a lower value for the grid efficiency is used (\eg, \verb|geffcy| $\simeq 0.7$) to try and balance minimising the number of refined zones with the overhead associated with managing an increasing number of small grids.

Evidently, \azeus is able to follow the blast wave closely, regardless of its orientation.  Table \ref{tab:blast} compares the extrema of the plotted variables between the AMR calculation and uniform grid solutions with $200^2$ and $400^2$ zones.  With the exception of the pressure, which exhibits differences between the uniform 400$^2$ and AMR solutions of $\lesssim$ 0.5\%, the uniform grid results bracket the AMR solution.

\begin{table}[t!]
  \begin{center}
  \begin{tabular}{ccccccc}
    \tableline
    & \multicolumn{2}{c}{AMR} & \multicolumn{2}{c}{Uniform, $200^2$} & \multicolumn{2}{c}{Uniform, $400^2$} \\
    & Min & Max & Min & Max & Min & Max \\
    \tableline\tableline
    $\rho$ & 0.199 & 3.36 & 0.200 & 3.22 & 0.189 & 3.40 \\
    $p$ & 0.712 & 32.3 & 0.771 & 32.0 & 0.714 & 32.1 \\
    $p_B$ & 24.3 & 75.7 & 24.9 & 76.0 & 23.6 & 75.6 \\
    \tableline
  \end{tabular}
  \end{center}
  \caption{\label{tab:blast} Extrema for density, $\rho$, gas pressure, $p$, and magnetic pressure, $p_B$, in AMR and uniform grid solutions of the 2-D MHD blast problem at $t = 0.02$.}
\end{table}

For this test, we set \verb|tolshk| = \verb|tolcd| = 0.2, and apply \verb|tolgrad| = 0.2 to $e_{\rm T}$.  While the gradient detector is useful in refining the initial pressure jump, most ($\sim$ 99.9\%) of the zones flagged for refinement soon thereafter are detected by the CD and shock detectors.  Additional parameter settings include \verb|kcheck| = 10, \verb|ibuff| = 3, $\gamma = 5/3$, ${\cal C} = 0.5$, \verb|qcon| = 1.0, and \verb|qlin| = 0.1.  All boundaries are set to outflow conditions.

\subsubsection{Orszag-Tang MHD vortex}
\label{ssub:otvortex}
The 2-D vortex problem of \citet{ot79} has become a standard test for astrophysical MHD codes, and as it has not previously been performed using our version of \zeusddd, we present both non-AMR and AMR results here.  It is important to note that the same code was used to produce both sets of results:  \azeus is designed to be modular, and by deselecting the `AMR' option at the precompilation step, the code reverts to \zeusddd.

For this test, we follow \citet{sgths08} by initialising a periodic, Cartesian box of size $x,y \in [0,1]$ with initially constant pressure and density, $P = 5/12\pi$ and $\rho = \gamma P = 25/36\pi$, for $\gamma = 5/3$.  The velocity is initialised to $(v_x, v_y, v_z) = ( - \sin\, (2\pi y),\, \sin\, (2\pi x),\, 0)$, and the magnetic field is set through the vector potential $A_z = (B_0/4\pi) \cos\, (4\pi x) + (B_0/2\pi) \cos\, (2\pi y)$, where $B_0 = 1/\sqrt{4\pi}$.  

\begin{figure}[t!]
  \begin{center}
    \includegraphics[width=1.0\textwidth]{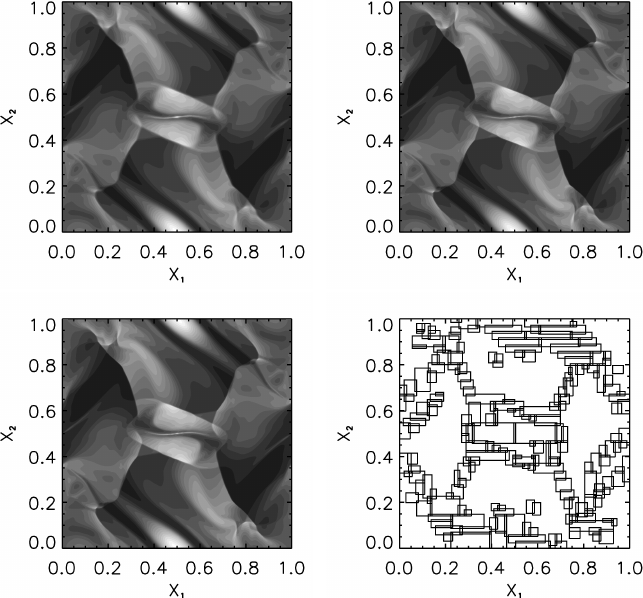}
  \end{center}
  \caption{\label{fig:ot} Uniform and adaptive grid solutions for the Orszag-Tang MHD vortex at $t = 1/2$.  Plotted are 20 evenly spaced contours of the gas pressure with range $[0.03, 0.50]$.  \textit{Top left:}  uniform grid solution with $256^2$ zones;  \textit{top right:} uniform grid solution with $512^2$ zones; \textit{bottom left:} AMR solution with a base grid resolution of $128^2$ and 2 levels of refinement; and \textit{bottom right:} distribution of grids at level 3 in the AMR solution.}
\end{figure}

The results for uniform grids with $256^2$ and $512^2$ zones at $t = 1/2$, as well as the AMR results for a base grid of $128^2$ zones and 2 levels of refinement (effective resolution of $512^2$ zones), are presented in Figure \ref{fig:ot}.  The bottom right panel shows the distribution of AMR grids at $t = 1/2$.  For clarity, we have only plotted the grids at level 3 (\ie, 2 levels higher resolution than the base grid). Even then, the filling factor of level $l = 2$ grids at this time is $\gtrsim 95\%$.  Examining the first three panels of Figure \ref{fig:ot} closely, features are noticeably sharper in the $512^2$ solution relative to the $256^2$ results, while the AMR and $512^2$ solutions are indistinguishable.

\begin{figure}[t!]
  \begin{center}
    \includegraphics[width=0.60\textwidth]{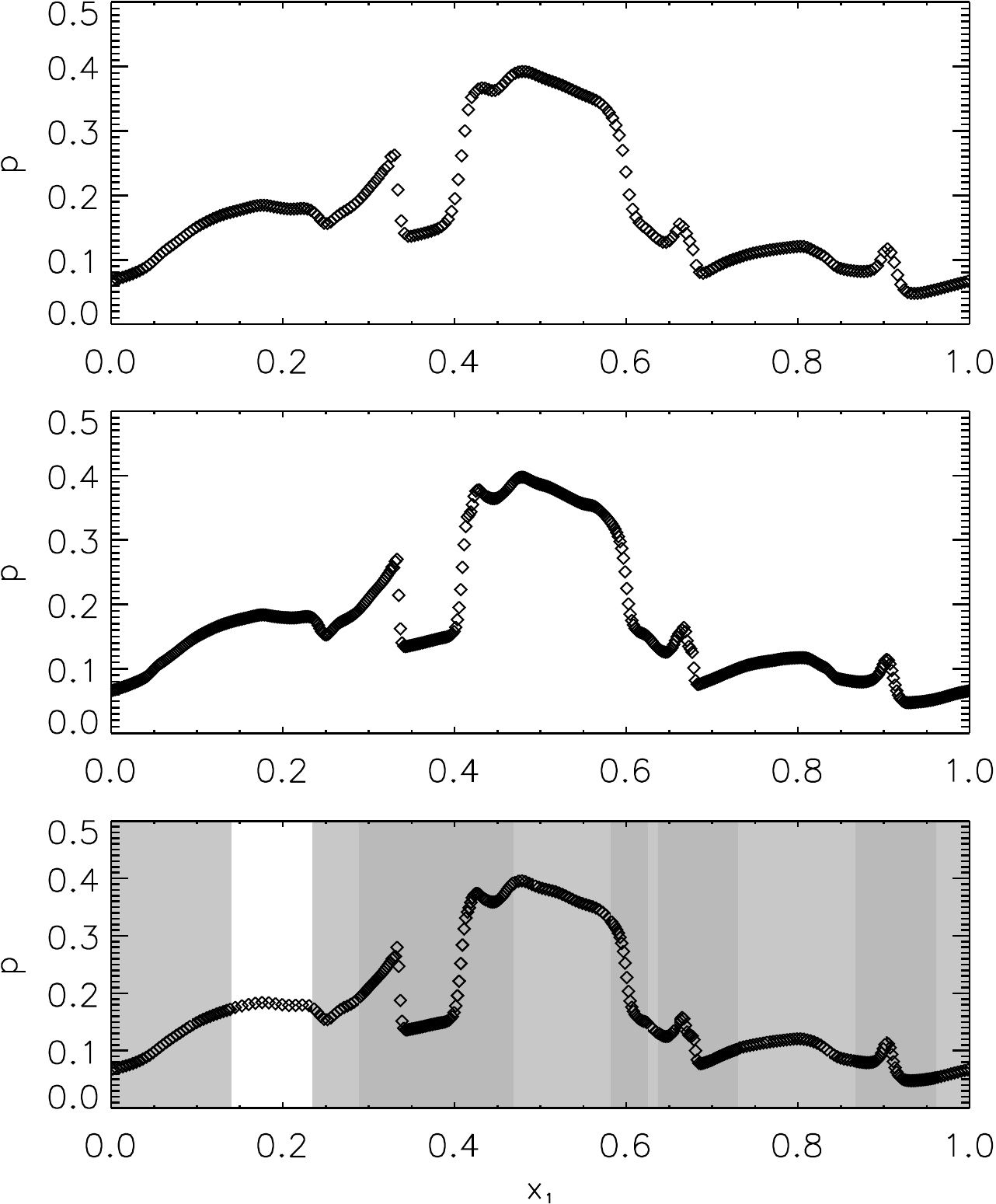}
  \end{center}
  \caption{\label{fig:otslice} 1-D slices of the gas pressure at $t = 1/2$ and $y = 0.4277$ in the Orszag-Tang MHD vortex problem.  From top to bottom, uniform $256^2$ grid, uniform $512^2$ grid, AMR solution with 2 levels of refinement and an effective resolution of $512^2$ zones.}
\end{figure}

Figure \ref{fig:otslice} shows slices of the gas pressure as a function of $x$ at $t = 1/2$ and $y = 0.4277$, which once again demonstrate that the AMR and $512^2$ uniform grid solutions are virtually identical.  Quantitatively, these solutions compare favourably with those from higher-order codes such as \textsl{ATHENA} \citep{sgths08}, with the discontinuities in the \azeus solutions being slightly broader.

For this problem, we set ${\cal C} = 0.5$, \verb|qcon| = 1.0, and \verb|qlin| = 0.1 for both uniform and adaptive grids.  For the AMR results, \verb|kcheck| = 10, \verb|geffcy| = 0.9, \verb|ibuff| = 2, and \verb|tolshk| = 0.2.  Neither the CD nor gradient detector were engaged.
\subsubsection{Magnetised accretion torus}
\label{ssub:torus}
This test is based on the simulations of \citet{h00} and \citet{m07} for a magnetised, constant angular momentum torus in axisymmetric spherical $(r, \vartheta)$ coordinates, and highlights the use of curvilinear coordinates in \azeus.

The torus structure is described by the equilibrium condition:
\begin{equation}
  \label{eq:torusequil}
  \frac{\gamma p}{\left(\gamma - 1\right) \rho} ~=~ C - \phi - \frac{1}{2}~ \frac{l^2_{\rm Kep}}{r^2 \sin^2 \!\vartheta},
\end{equation}
where $C$ is a constant of integration, $\phi = - 1 / (r - 1)$ is the pseudo-Newtonian gravitational potential, and $l_{\rm Kep}$ is the Keplerian angular momentum at the pressure maximum.  The pressure $p$ is initially related to the density via the polytropic relation, $p = \kappa \rho^{\gamma}$, with $\gamma = 5/3$.  By specifying the location of the pressure maximum ($r_{\rm max} = 4.7$) and the inner edge of the torus ($r_{\rm min} = 3$), we can determine the value of $C$ and the (hydrodynamic) structure of the torus.

A poloidal magnetic field is initialised in the torus from the $\varphi$-component of the vector potential:
\begin{equation*}
A_\varphi ~=~ \frac{B_0}{\rho_{\rm m}}\, {\rm min}\big( \rho(r,\vartheta) - \rho_{\rm c}, 0\big),
\end{equation*}
where $B^2_0 = 2\kappa\rho_{\rm m}^\gamma / \beta_{\rm m}$, $\beta_{\rm m}$ and $\rho_{\rm m}$ are, respectively, the plasma-beta and density at $r_{\rm max}$, $\kappa$ is determined by equation (\ref{eq:torusequil}) evaluated at $r_{\max}$, and $\rho_{\rm c} = \rho_{\rm m}/2$ determines the surface of the last vector equipotential.  For the simulations presented here, $\beta_{\rm m}= 350$ and $\rho_{\rm m} = 10$.  The toroidal velocity ($v_\varphi$) in the torus is initialised to the local Keplerian speed, while the poloidal velocity is set to zero everywhere.

Outside the torus, the magnetic field is zero and we initialise a hydrostatic atmosphere with density and temperature contrasts of $\rho_{\rm atm} / \rho_{\rm m} = 10^{-4}$ and $T_{\rm atm} / T_{\rm m} = 100$, respectively, where $T_{\rm m}$ is the temperature at $r_{\rm max}$.

The domain of the grid is $r\in[1.5,20]$ and $\vartheta\in[0,\pi/2]$.  We impose reflecting boundary conditions suitable for a rotation axis at $\vartheta = 0$, reflecting/conducting boundary conditions at the equatorial plane ($\vartheta = \pi/2$), and outflow boundary conditions at $r=20$.  At $r=r_{\rm in}=1.5$, we impose ``sink'' boundary conditions in an attempt to absorb any material reaching the inner boundary.  This involves maintaining the density and pressure at their initial values, and setting $v_r = v_\vartheta = v_\varphi =0$ within the boundary.  On the inner skin ($r=r_{\rm in}$), $v_r$ is set to the minimum of zero and a linear extrapolation from the grid.  Finally, outflow boundary conditions are applied to the EMFs:
\begin{align}
  \label{eq:emfbcstorus}
  \cale_1(r < r_{\rm in}) & ~=~ \cale_1(2r_{\rm in}-r); \nonumber \\
  \cale_2(r < r_{\rm in}) & ~=~ 2\cale_2(r_{\rm in}) - \cale_2(2r_{\rm in}-r); \\
  \cale_3(r < r_{\rm in}) & ~=~ 2\cale_3(r_{\rm in}) - \cale_3(2r_{\rm in}-r). \nonumber
\end{align}
Values of $\cale_2(r_{\rm in})$ and $\cale_3(r_{\rm in})$ are determined by the CMoC algorithm in \azeus.

Both the uniform grid and AMR solutions are presented here. For the single grid calculation, we use 592 uniform radial zones and 256 uniform meridional zones. For the AMR solution, we use a base grid of $296 \times 128$ zones, and allow one level of refinement with a refinement ratio $\nu = 2$, giving the same effective resolution as the uniform grid calculation.

\begin{figure}[t!]
  \begin{center}
    \includegraphics[width=1.0\textwidth]{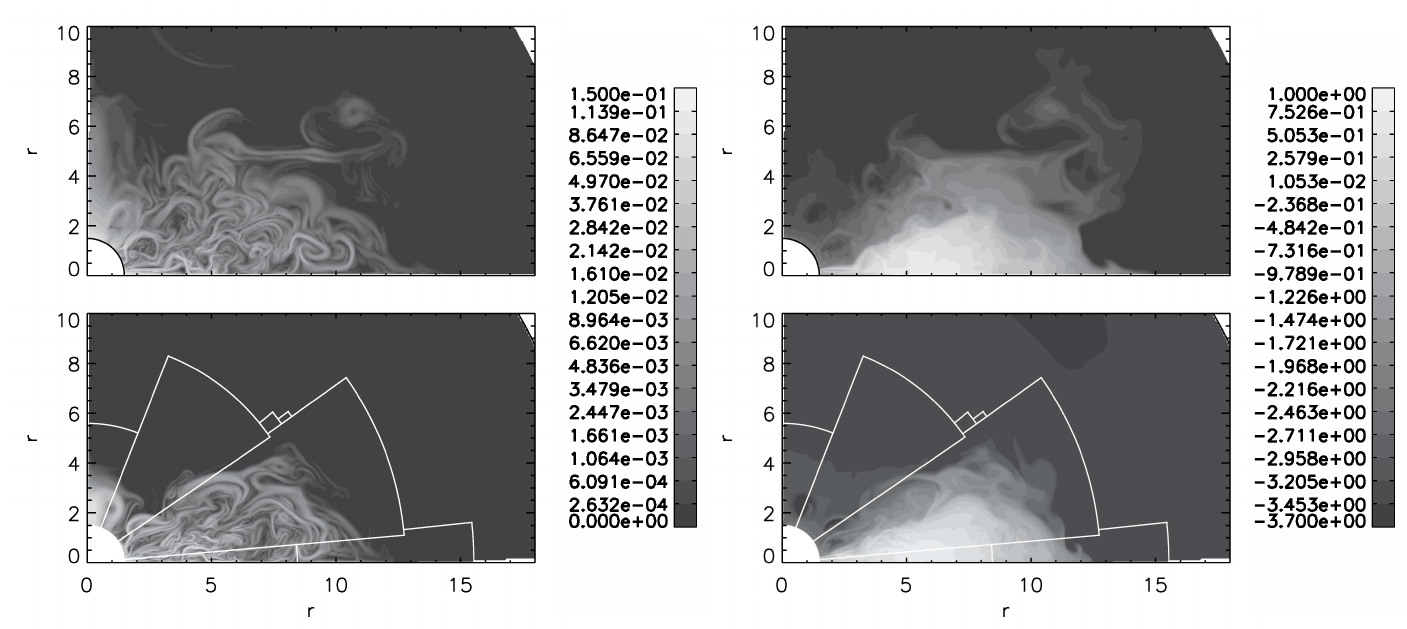}
  \end{center}
  \caption{\label{fig:torus} The ``beard of \textsf{AZEuS}": Results of MHD accretion torus simulations in $(r, \vartheta)$ coordinates at $t = 300$, where the vertical axis is the rotation axis.  Plotted are contours of poloidal magnetic field (left) and logarithmic density (right).  Top panels are for the uniform grid, bottom panels the AMR solution with one level of refinement.  Borders of the adaptive grids are shown with white lines.}
\end{figure}

For both calculations, ${\cal C}$ = 0.5, \verb|qcon| = 1.0, and \verb|qlin| = 0.1.  For AMR, \verb|kcheck| = 50, \verb|geffcy| = 0.8, \verb|ibuff| = 2, \verb|tolshk| = \verb|tolcd| = 0.2, and \verb|tolgrad| = 0.2 is applied to $B_\vartheta$.  We use the gradient detector to refine on the initial magnetic field configuration (contained entirely within the torus), and the complex field structure that develops later on from the magneto-rotational instability (MRI; \citealt{bh91}), since neither are well-tracked by the shock and CD detectors.

Following \citet{h00}, we enforce a density floor of $10^{-3} \rho(r_{\rm in})$ to prevent the time step from becoming prohibitively small, and we use the internal energy equation to avoid negative pressures.  Unlike \citet{h00}, we do not introduce any perturbations on the initial conditions, which slows but does not prevent the onset of MRI.

Figure \ref{fig:torus} presents the results of our simulations at $t = 300$.   Evidently and very much unlike the Orszag-Tang vortex, the AMR and uniform grid solutions are visually different, even though they have the same effective resolution.  For a pseudo-turbulent, ``irreversible" problem such as this, small differences between values in the coarse and fine grids grow exponentially during the simulation, and lead to visually different solutions.  This is quite unlike the behaviour of a ``reversible" problem such as the Orszag-Tang vortex, for which small fluctuations grow at worse linearly in time, and never manifest as visual differences in the plots.  On the other hand, integrated quantities tend to be in better agreement than their detailed distributions.  For example, between the uniform grid and AMR solutions, the integrated mass and kinetic energy at $t = 300$ differ by only 0.9\% and 6.5\% respectively.  Furthermore, our results are qualitatively similar to \citet{m07}; Figure \ref{fig:torus} could easily fit in with their Figure 8 for different Riemann solvers and interpolation schemes.

\begin{figure}[t!]
  \begin{center}
    \includegraphics[width=0.9\textwidth]{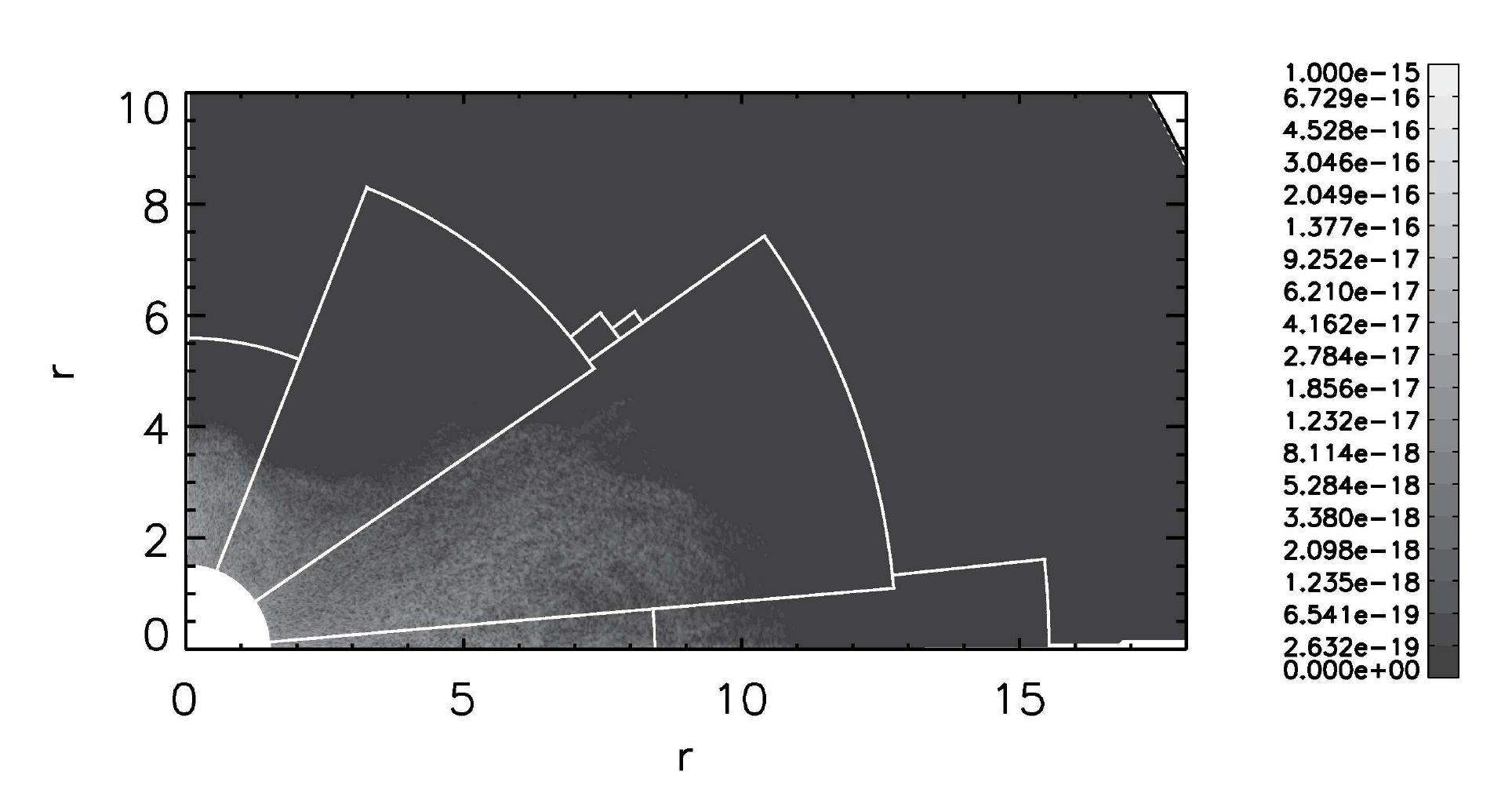}
  \end{center}
  \caption{\label{fig:divb} \azeus results for the normalised magnetic field divergence, $\big|\nabla\cdot\vec{B}\big|\, \Delta x / |\vec{B}|$, in the MHD accretion torus simulation at $t = 300$.  The maximum value of the normalised divergence at this time, including both physical and adjacent boundary zones, is $1.34 \times 10^{-15}$.  Borders of the adaptive grids are shown with white lines.}
\end{figure}

Figure \ref{fig:divb} shows the normalised magnetic field divergence, $\big|\nabla\cdot\vec{B}\big|\, \Delta x / |\vec{B}|$, at the simulation end time, where the maximum value is $1.34 \times 10^{-15}$, including both physical and adjacent boundary zones.  Indeed, the minimum and maximum values for the divergence throughout the simulation life time reach $-7.392\times 10^{-15}$ and $7.396\times 10^{-15}$ respectively, demonstrating that the algorithms employed for prolongation and restriction of the magnetic field in \azeus are capable of maintaining the solenoidal condition to machine round-off, assuming the initial conditions are also divergence-free.  It is also worthwhile to note the absence of grid boundary effects in Figure \ref{fig:divb}, which one might expect if monopoles existed within the boundaries of the fine grids.

\subsection{A 3-D test: self-gravitational hydrodynamical collapse}
\label{sub:truelove}
The last test is the self-gravitational hydrodynamical collapse calculation first presented by \citet{t97} and \citet[\citetalias{t98}]{t98}.  At the time of this writing, self-gravity has not been fully implemented in \azeus, and so we use the successive over-relaxation (SOR) method \citep{p92} as an easy-to-program, albeit slow, stop-gap measure. Since SOR is incompatible with periodic boundary conditions, we apply inflow boundary conditions instead.  This results in some subtle differences from \citetalias{t98}, which we discuss below.

A rotating, uniform cloud of mass $M = 1 M_\odot$ and radius $R = 5 \times 10^{16}$ cm is initialised in the centre of a 3-D Cartesian box with side length $4R$.  We use a nearly isothermal equation of state ($p = \kappa\rho^\gamma$, $\gamma = 1.001$), initially uniform rotation $\Omega = 7.14 \times 10^{-13}$ rad s$^{-1}$ with angular momentum axis in the positive $x_3$-direction, and energy ratios of
\begin{equation}
  \label{eq:trueratios}
   \alpha ~=~ \frac{5}{2} \left(\frac{3}{4\pi\rho_0 M^2}\right)^{1/3} \frac{c_{\rm s}^2}{G} ~=~ 0.16 \quad \textrm{and} \quad \beta_\Omega ~=~ \frac{1}{4\pi} \frac{\Omega^2}{G\rho_0} ~=~ 0.26,
\end{equation}
where $\rho_0$ and $c_{\rm s}$ are, respectively, the initial density and sound speed of the uniform cloud \citepalias{t98}.  The remainder of the computational volume is initialised with uniform density and pressure given by $\rho_{\rm atm} = 100 \rho_{\rm cloud}$ and $p_{\rm atm} = 10 p_{\rm cloud}$.  Upon the initial uniform density distribution, we apply an azimuthal $m = 2$ perturbation of the form $\rho_{\rm cloud} \big(1 + A \cos (2\varphi)\big)$, where $A = 10\%$ is the perturbation amplitude and $\varphi$ is the azimuthal angle relative to the cloud centre.

Following \citetalias{t98}, we set the coarsest grid to $32^3$ zones and designate it $R_8$ (meaning the cloud radius is resolved with eight zones).  We immediately add 1 level of refinement by flagging all zones which have density $\rho > \rho_{\rm cloud}$ (for a refinement ratio of 4, this gives 32 zones per cloud radius, or $R_{32}$).  Beyond this, we enforce the so-called ``Truelove criterion'':
\begin{equation}
  \label{eq:truecrit}
  J ~=~ \frac{\Delta x}{\lambda_J} ~=~ \Delta x \left(\frac{G\rho}{\pi c_{\rm s}^2}\right)^{1/2} <~ J_{\rm max},
\end{equation}
refining zones which have Jeans numbers $J$ larger than critical value $J_{\rm max} = 0.25$.  Additional run-time parameters for this test include ${\cal C} = 0.33$, \verb|qcon| = 1.0, \verb|qlin| = 0.1, \verb|kcheck| = 20, \verb|geffcy| = 0.7, \verb|ibuff| = 2, and refinement ratio $\nu = 4$.

\begin{figure}[t!]
  \begin{center}
    \includegraphics[width=1.0\textwidth]{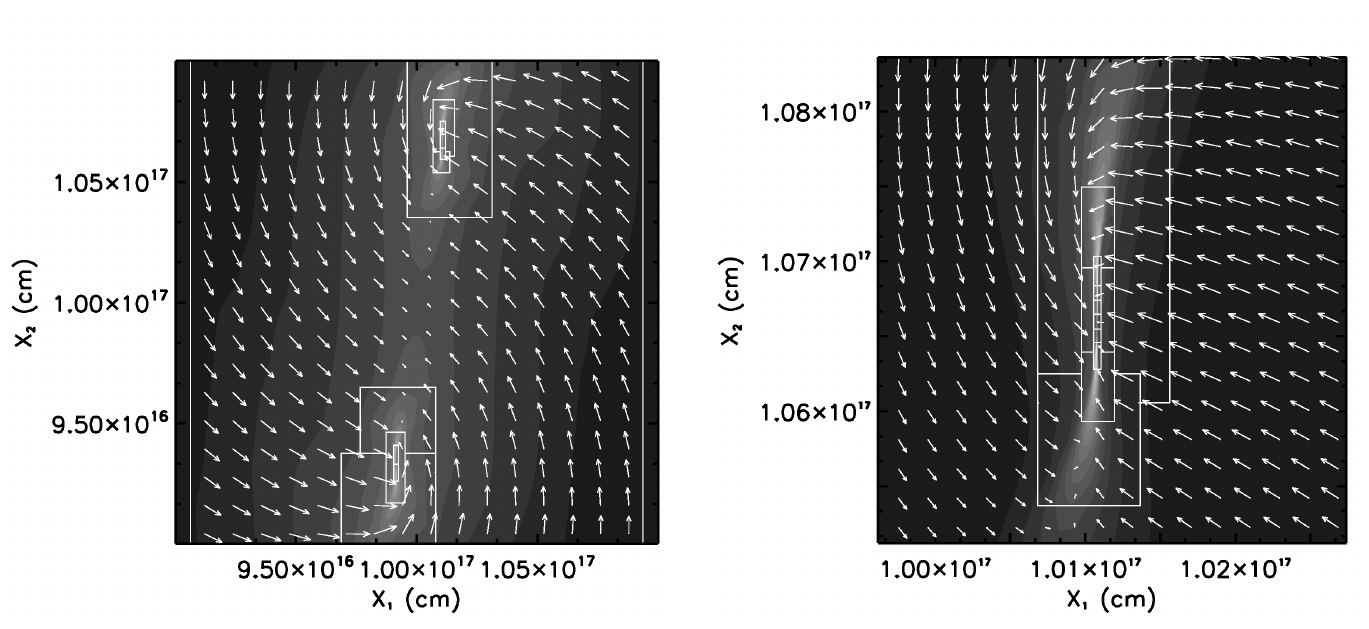}
  \end{center}
  \caption{\label{fig:true} \azeus results for the Truelove problem with a 10\% amplitude perturbation at $t = 0.59808 = 1.2152t_{\rm ff}$, where $t_{\rm ff}$ is the free-fall time.  \textit{Left:} Equatorial $x_1$-$x_2$ slice of logarithmic density with velocity vectors superposed.  The highest resolution shown here is $R_{8192}$, five levels of refinement above the base grid. \textit{Right:} Equatorial $x_1$-$x_2$ slice of logarithmic density of the upper fragment.  The highest resolution in this plot is $R_{32768}$ (six levels) with one additional level of refinement not shown.  White lines denote the borders of AMR grids and the units of density are g cm$^{-3}$.  For each panel, 20 evenly spaced contours are plotted with ranges $\log \rho = [-16.10, -9.905]$ (left) and $[-14.30, -9.604]$ (right).}
\end{figure}

Figure \ref{fig:true} shows our results at $t = 0.59808 = 1.2152t_{\rm ff}$.  At this time, there are seven levels of refinement above the base grid ($R_{131072}$).  Our maximum density at this time is $\log \rho_{\rm max} = -9.347399$, with density measured in g cm$^{-3}$, an increase over $\rho_0$ of more than 8 orders of magnitude.  While our simulation collapses more quickly than \citetalias{t98} and our Figure \ref{fig:true} does not correspond exactly with their Figures 12 and 13, we also reach a somewhat lower maximum density.

The differences between our results and \citetalias{t98} are most likely caused by the differing boundary conditions (ours inflow, theirs periodic).  Indeed, \citetalias{t98} allude to this possibility, suggesting that non-periodic boundary conditions could slightly increase the rate of collapse, which we observe.
\section{Summary}
\label{sec:summary}
We have described a method for block-based AMR on a fully-staggered mesh, and implemented this method in a new version of \zeusddd called \azeus.  In addition to describing the modifications required to AMR to account for the fully staggered grid, we also describe higher order interpolation methods for the prolongation step which we found necessary to allow for static grids.  Static grids are important for problems which, at first order, have a self-similar character and expand over the course of the simulation to ever larger scale lengths.  Such a simulation by \azeus has already appeared in the literature \citep{rc11}, which showcases the ability of the code to transmit waves of all types and strengths across grid boundaries, and to do so in cylindrical coordinates.  The higher-order prolongation operator is designed to maintain the conservation of all important physical quantities such as mass, momentum, energy, and magnetic flux.  

Numerous test problems were also presented in 1-, 2-, and 3-D, and in both Cartesian and spherical polar coordinates.  These tests demonstrate the ability of the code to produce essentially identical results in ``reversible" (non-turbulent) problems whether using a single grid or AMR, and give an example of the differences that can occur in ``irreversible" (turbulent) problems as minute differences caused by the insertion or deletion of a grid amplify. 

Finally, the \azeus website \href{http://www.ica.smu.ca/azeus}{\texttt{http://www.ica.smu.ca/azeus}} was introduced on which test problems and simulations will be posted as they become available, and from which the code can be downloaded in the near future.
\acknowledgements{We thank the referee for a careful reading of the manuscript and for pointing out an oversight in the algorithm as described in the original draft.  We also thank Marsha Berger for providing her AMR subroutines.  This work is supported, in part, by the Natural Sciences and Engineering Research Council (NSERC).  Computing resources were provided by ACEnet, which is funded by CFI, ACOA, and the provinces of Nova Scotia, Newfoundland \& Labrador, and New Brunswick.}
\appendix
\section{Curvilinear Coordinates}
\label{app:curvi}
Like other codes in the \zeus family, \azeus uses the ``covariant" (coordinate independent) form of the MHD equations \citep{sn92a}.  Supported geometries in \azeus include Cartesian ($x, y, z$), cylindrical ($z, r, \varphi$), and spherical polar ($r, \vartheta, \varphi$) coordinates while other orthogonal curvilinear coordinate systems may be easily implemented as needed.  For simplicity, the equations presented in the main body of the paper were all written assuming Cartesian-like coordinates.  In this appendix, we show how these may be rewritten in a covariant fashion.

The distance differential in an orthogonal coordinate system, $(x_1, x_2, x_3)$, is given by:
\begin{equation}
  \label{eq:metric}
  ds^2 = g_1^2\,dx_1^2 + g_2^2\,dx_2^2 + g_3\,dx_3^2,
\end{equation}
where $g_i$, $i=1,2,3$, are the usual metric scaling factors, all functions of the coordinates. If the scaling factors are separable and independent of their own direction, we can write:
\begin{alignat}{3}
  \label{eq:seperable}
  g_1 ~=~ g_1(j,k) & ~=~ g_{12}(j)\,g_{13}(k);& \nonumber\\
  g_2 ~=~ g_2(k,i) & ~=~ g_{23}(k)\,g_{21}(i);& \\
  g_3 ~=~ g_3(i,j) & ~=~ g_{31}(i)\,g_{32}(j),& \nonumber
\end{alignat}
listed in Table \ref{tab:metrics} for Cartesian, cylindrical, and spherical polar coordinates.  
\begin{table}[t!]
  \begin{center}
  \begin{tabular}{c|cccccc}
   $(x_1, x_2, x_3)$ & $g_{12}(j)$ & $g_{13}(k)$ & $g_{23}(k)$ & $g_{21}(i)$ & $g_{31}(i)$ & $g_{32}(j)$ \\
   \tableline\tableline
   $(x, y, z)$       & 1 & $1$ & 1 & 1 & 1 & 1 \\
   $(z, r, \varphi)$ & 1 & 1 & 1 & 1 & 1 & $x_2(j)$ \\
   $(r, \vartheta, \varphi)$ & 1 & 1 & 1 & $x_1(i)$ & $x_1(i)$ & $\sin x_2(j)$ \\
  \end{tabular}
  \caption{\label{tab:metrics} Metric scaling factors in \azeus for Cartesian, cylindrical, and polar coordinates.}
  \end{center}
\end{table}

Face areas and zone volumes important for the calculation of fluxes and conserved quantities are written as:
\begin{align}
  \label{eq:darea1}
  \delta A_1(i,j,k) & ~=\, \int_{x_2(j)}^{x_2(j+1)}\!\!\int_{x_3(k)}^{x_3(k+1)}\!\!\! g_{21}(i)\,g_{31}(i)\,g_{32}(j')\,  dx_2'\, dx_3'\nonumber\\
  & ~=~ g_{21}(i)\,g_{31}(i)\! \int_{x_2(j)}^{x_2(j+1)}\!\!\! g_{32}(j')\,dx_2'\! \int_{x_3(k)}^{x_3(k+1)}\!\!\! dx_3'\\
  & ~\equiv~ g_{21}(i)\,g_{31}(i) \,\delta A_{12}(j)\, \delta A_{13}(k)\nonumber,\\[6pt]
  \label{eq:darea2}
  \delta A_2(i,j,k) & ~=\, \int_{x_3(k)}^{x_3(k+1)}\!\!\int_{x_1(i)}^{x_1(i+1)}\!\!\! g_{31}(i')\, g_{32}(j)\, dx_3'\, dx_1'\nonumber\\
  & ~=~ g_{32}(j)\!\int_{x_3(k)}^{x_3(k+1)}\!\!\! dx_3'\! \int_{x_1(i)}^{x_1(i+1)}\!\!\! g_{31}(i')\,dx_1'\\
  & ~\equiv~ g_{32}(j)\, \delta A_{23}(k)\, \delta A_{21}(i)\nonumber,\\[6pt]
  \label{eq:darea3}
  \delta A_3(i,j,k) & ~=\, \int_{x_1(i)}^{x_1(i+1)}\!\!\int_{x_2(j)}^{x_2(j+1)}\!\!\! g_{21}(i')\,dx_1'\,  dx_2'\nonumber\\
  & ~=\, \int_{x_1(i)}^{x_1(i+1)}\!\!\! g_{21}(i')\,dx_1' \!\int_{x_2(j)}^{x_2(j+1)}\!\!\! dx_2'\\
  & ~\equiv~ \delta A_{31}(i) \, \delta A_{32}(j).\nonumber
\end{align}
\begin{align}
  \label{eq:dvolume}
  \delta V(i,j,k) & ~=\, \int_{x_1(i)}^{x_1(i+1)}\!\!\int_{x_2(j)}^{x_2(j+1)}\!\!\int_{x_3(k)}^{x_3(k+1)}\! \! g_{21}(i')\, g_{31}(i')\, g_{32}(j')\, dx_1'\, dx_2'\, dx_3'\nonumber\\
                  & ~=\, \int_{x_1(i)}^{x_1(i+1)}\!\!\! g_{21}(i')\, g_{31}(i')\, dx_1' \!\int_{x_2(j)}^{x_2(j+1)}\!\!\! g_{32}(j')\, dx_2' \!\int_{x_3(k)}^{x_3(k+1)}\!\!\! dx_3' \\
                  & ~\equiv\, \int_{x_1(i)}^{x_1(i+1)}\!\!\! dV_1' \int_{x_2(j)}^{x_2(j+1)}\!\!\! dV_2' \int_{x_3(k)}^{x_3(k+1)}\!\!\! dV_3' \\
                  & ~\equiv~ \delta V_1(i)\, \delta V_2(j)\, \delta V_3(k)\nonumber,
\end{align}
where metric scaling factors equal to 1 in Table \ref{tab:metrics} have been and will continue to be dropped.  

Finally, \zeus has traditionally defined the momentum densities, $\vec s$, as:
\begin{align}
  \label{eq:curvmom}
  s_1(i,j,k) & ~=~ \,\rho(i-\tfrac{1}{2},j,k)\, v_1(i,j,k);\nonumber\\
  s_2(i,j,k) & ~=~ \,g_{21}(i)\,\rho(i,j-\tfrac{1}{2},k)\, v_2(i,j,k);\\
  s_3(i,j,k) & ~=~ \,g_{31}(i)\,g_{32}(j)\,\rho(i,j,k-\tfrac{1}{2})\, v_3(i,j,k),\nonumber
\end{align}
where the half indices in the density indicate two-point averages.  Defining the momentum components with the metric scaling factors simplifies the momentum equation somewhat by eliminating all ``Coriolis-like" fictitious forces, leaving only the ``centrifugal-like" terms.
\subsection{Restriction}
\label{appsub:restrict}
\subsubsection{Conservative Overwrite}
\label{appssub:overwrite}
When applying the conservative overwriting procedure on a curvilinear grid, the equations of Section \ref{sub:overwrite} must be modified to account for the non-constant volumes of zones.  For example, equation (\ref{eq:over_zcvc}) for zone-centred quantities becomes:
\begin{align}
  \label{eq:over_zcvc_curve}
  Q(I,J,K)\, \Delta V_1(I)\,\Delta V_2(J)\,\Delta V_3(K) ~= & \sum_{\alpha, \beta, \eta = 0}^{\nu-1}\, q(i+\alpha, j+\beta, k+\eta)\\
  &~~\times\,\delta V_1(i+\alpha)\,\delta V_2(j+\beta)\,\delta V_3(k+\eta),\nonumber
\end{align}
and equation (\ref{eq:over_fcvc}) for face-centred momenta generalises to:
\begin{align}
  \label{eq:over_fcvc_curve}
  S_1(I,J,K)\,\Delta V_1(I)\,\Delta V_2(J)\,\Delta V_3(K) ~= \sum_{\alpha=-\nu/2}^{\nu/2}& \, \sum_{\beta, \eta = 0}^{\nu-1}\, {\cal G}'(\alpha)\, s_1(i+\alpha, j+\beta, k+\eta)\\ 
  &\times\delta V_1(i+\alpha)\,\delta V_2(j+\beta)\,\delta V_3(k+\eta),\nonumber
\end{align}
where:
\begin{equation}
  \label{eq:weighting}
  \quad {\cal G}'(\alpha) ~=~
  \begin{cases}
    w_{\rm 1,R}(i + \alpha) & \textrm{if } \alpha = -\nu/2; \\
    w_{\rm 1,L}(i + \alpha) & \textrm{if } \alpha = +\nu/2; \\
    1 & \textrm{otherwise}, 
  \end{cases}
\end{equation}
and where:
\begin{equation}
  \label{eq:weightfact}
  w_{\rm 1,L}(i) ~=~ \frac{1}{\delta V_1(i)} \int_{x_1(i-1/2)}^{x_1(i)} \!dV_1; \quad w_{\rm 1,R}(i) ~=~ \frac{1}{\delta V_1(i)} \int_{x_1(i)}^{x_1(i+1/2)} \!dV_1.
\end{equation}
While it is still true that the fine zones at $\alpha = \pm \nu/2$ are halfway outside the coarse ROI in \emph{position} space (Figure \ref{fig:rois}b), this is not generally true in volume space.  As an example, consider spherical coordinates where the 1-volume differential is $dV_1(i) = g_{21}(i) g_{31}(i)\, dx_1(i) = x_1^2(i)\, dx_1(i)$.  Clearly, $dV_1(i)$ in spherical coordinates is not a linear function of position, and it cannot be assumed that exactly 1/2 of the fine momentum volume is outside the coarse ROI.  To correct for this, we calculate the ratio of the \emph{half}-volume inside the coarse ROI to the actual volume element (equation \ref{eq:weightfact}), which is then used as a weighting factor in ${\cal G}'(\alpha)$.  For reasonable grid parameters, these weighting factors are small corrections ($<$ a few \%) relative to the Cartesian factor of $1/2$, but nonetheless are included for accuracy.

Suitable expressions for the 2-direction are obtained by a permutation of indices.  No corrections are necessary for the 3-direction, since $dV_3(k) = dx_3(k)$ for the curvilinear coordinates discussed here.

For the magnetic field, equation (\ref{eq:over_fcac}) changes to account for non-uniform areas:
\begin{equation}
  \label{eq:over_fcac_curve}
   B_1(I,J,K)\,\Delta A_1(I,J,K) ~=\, \sum_{\beta, \eta = 0}^{\nu-1}\, b_1(i, j+\beta, k+\eta)\,\delta A_1(i,j+\beta,k+\eta).
\end{equation}
\subsubsection{Flux Corrections}
\label{appssub:reflux}
Fluxes and EMFs as stored by \zeusddd already include the appropriate area and length elements, so the required changes for flux corrections in curvilinear coordinates are not substantial.  As with the conservative overwrite, the corrections (\eg, equation \ref{eq:case2s2}) need to be adjusted by replacing occurrences of ${\cal G}$ (equation \ref{eq:gfactor}) with ${\cal G}'$ (equation \ref{eq:weighting}).  Similar to equation (\ref{eq:over_fcvc_curve}), equation (\ref{eq:case1s1}) for the flux corrections of momentum components normal to the boundary must also be modified to account for non-constant volumes:
\begin{equation}
  \label{eq:rfx_s1_curv}
  \begin{split}
    \widetilde{S}_1^{N+1}(I,J,K) ~=~ S_1^{N+1}(I,J,K) \, -\,& \frac{1}{\Delta V(I,J,K)}\bigg[ F_{1,S_1}^{N+\onehalf}(I-1,J,K) \\[3pt]
   - \sum_{\beta,\eta,\tau=0}^{\nu-1} \Big(w_{\rm 1,R}&\big(i-\tfrac{\nu}{2}\big)\,f_{1,s_1}^{n+\tau+\onehalf}\big(i-\tfrac{\nu}{2}-1, j+\beta, k+\eta\big) \\[-3pt]
    +w_{\rm 1,L}&\big(i-\tfrac{\nu}{2}\big)\,f_{1,s_1}^{n+\tau+\onehalf}\big(i-\tfrac{\nu}{2}, j+\beta, k+\eta\big) \Big) \bigg].
  \end{split}
\end{equation}

By the same argument, equation (\ref{eq:case3s2}) for coarse momenta with flux components parallel to a grid boundary also needs to be modified, as the original factor of $(\nu - 1) / 2\nu \equiv {\cal R}$ is not correct for curvilinear coordinates:
\begin{align}
\notag
  \widetilde{S}_2^{N+1}(I,J,K) ~&=~ S_2^{N+1}(I,J,K)\\
  \label{eq:flux_fcvc_curve}
&- \frac{1}{\Delta V(I,J,K)} \Bigg[\, {\cal L}(J,\nu)\bigg( F_{1,S_2}^{N+\onehalf}(I,J,K)-F_{1, S_2}^{N+\onehalf}(I+1,J,K)\bigg) \\
\notag
&\quad - \sum_{\beta=1}^{\nu/2}\, \sum_{\eta,\tau=0}^{\nu-1}\, {\cal G}'(\beta)\bigg( f_{1,s_2}^{n+\tau+\onehalf}(i,j+\beta,k+\eta) - f_{1,s_2}^{n+\tau+\onehalf}(i+\nu,j+\beta,k+\eta) \bigg)\Bigg].
\end{align}
Here, ${\cal L}(J,\nu) = W_{2,L}(J,\nu)$ if $S_2(I,J,K)$ is at the upper edge of the fine grid (high $j$), and ${\cal L}(J,\nu) =  W_{2,R}(J,\nu)$ if $S_2(I,J,K)$ is at the lower edge of the fine grid (low $j$), with:
\begin{equation*}
  W_{2,L}(J,\nu) ~=~ \frac{1}{\Delta V_2(J)} \int_{x_2(J-1/2)}^{x_2(J-1/2+{\cal R})} \!dV_2;
\end{equation*}
\begin{equation*}
  W_{2,R}(J,\nu) ~=~ \frac{1}{\Delta V_2(J)} \int_{x_2(J+1/2-{\cal R})}^{x_2(J+1/2)} \!dV_2.
\end{equation*}
Like the function ${\cal G}'$, ${\cal L}$ accounts for the portion of the coarse zone flux in volume space which is being replaced by fine fluxes.  As before, expressions for the 2-direction are obtained by a suitable permutations of indices.
\subsection{Prolongation}
\label{appsub:prolong}
The changes required for prolongation on curvilinear grids deal entirely with making the interpolations conservative for curvilinear volumes and areas.  In the case of 1-D PPI, equation (\ref{eq:ppiconserve}) summarising the conservation constraint is modified to:
\begin{equation}
  \label{eq:ppiconserve_curve}
  Q(I,J,K) \Delta V_1(I) ~=~ \sum^{\nu-1}_{\alpha=0}\, q(i+\alpha, j, k) \delta V_1(i),
\end{equation}
which then affects the rest of the scheme (equations \ref{eq:1dppi} and \ref{eq:ppifuncs}):
\begin{equation}
  \label{eq:1dppi_curve}
    q^*_1(i+\alpha) ~=~ Q_L(I) + \zeta \Big( Q_R(I) - Q_L(I) + {\cal H}_1' \big( 1 - \zeta \big) \Big),
\end{equation}
where:
\begin{align}
  \zeta &~=~ \frac{x_1(i+\alpha) - x_1(I)}{\Delta x_1(I)};\nonumber \\
 {\cal H}_1' & ~=~ \frac{1}{f_1'(\nu,i) - f_2'(\nu,i)} \Big(\Delta V_1(I) \big(Q(I,J,K) - Q_L(I)\big) - f_1'(\nu,i) \big( Q_R(I) - Q_L(I) \big)\Big);  \nonumber \\
  f_1'(\nu,i) & ~=~ \frac{1}{2\nu} \sum_{\xi=1}^{\nu} \big(2\xi - 1\big)\delta V_1(i+\xi-1); \nonumber \\
  f_2'(\nu,i) & ~=~ \frac{1}{4\nu^2} \sum_{\xi=1}^{\nu} \big(2\xi - 1\big)^2\delta V_1(i+\xi-1). \nonumber
\end{align}

As for PLI, the original 1-D scheme in the 1-direction takes the form \citep{vL77}:
\begin{equation}
  \label{eq:1dpli}
  q^*_1(i+\alpha)  ~=~ Q(I) + \big(2\zeta-1\big)\,\Delta Q',
\end{equation}
where:
\begin{equation*}
  \Delta Q' ~=~ \left\{{\begin{array}{ll} \dfrac{\Delta Q_{\rm R}\, \Delta Q_{\rm L}}{\Delta Q_{\rm R} + \Delta Q_{\rm L}} & {\textrm{if } \Delta Q_{\rm R}\, \Delta Q_{\rm L} > 0}; \\[12pt] {0,} & {\textrm{otherwise}},\end{array}}\right.
\end{equation*}
and where:
\begin{equation*}
\Delta Q_{\rm R} ~=~ Q(I+1)-Q(I); \quad \Delta Q_{\rm L} ~=~ Q(I)-Q(I-1).
\end{equation*}

Equation (\ref{eq:1dpli}) will not generally conserve mass, momentum, or magnetic flux on a curvilinear grid.  To correct this, we relax the constraint that the linear interpolation profile must pass through $Q(I)$ at the zone-centre.  This releases one degree of freedom that can then be used with equation (\ref{eq:ppiconserve_curve}), resulting in:
\begin{equation}
  \label{eq:1dpli_curve}
  q^*_1(i+\alpha) ~=~ Q(I) + \Delta Q'\left(\, 2\zeta-1 - \frac{1}{\Delta V_1(I)}\,f_3'(\nu,i)\right),
\end{equation}
where:
\begin{equation*}
f_3'(\nu,i)  ~=~ \frac{1}{\nu}\,\sum_{\xi=1}^{\nu} \,\big(\,(2\xi - 1) - \nu\,\big)\,\delta V_1(i+\xi-1).\nonumber
\end{equation*}
The additional term in equation (\ref{eq:1dpli_curve}) can be viewed as either a shift in the zone-centred intercept, or as choosing a modified value of $\zeta$ corresponding to the volume-centred rather than the spatial-centred coordinate.

Unfortunately, this ``shift'' can push $q^*_1(i+\alpha)$ beyond neighbouring values of $Q(I)$ at the edges of the interpolation profile, resulting in a loss of monotonicity.  However, since $\zeta = 0$ or $1$ at the left or right side of the interpolation profile, we can write:
\begin{equation}
  \label{eq:1dpli_mono1}
  \begin{split}
    \Delta Q_{\rm L}' & ~=~ Q(I) - q^*_1(i+\alpha) ~=~ \Delta Q' \big(\,1 + {\cal K}_1'\,\big) \\[6pt]
    \textrm{or}\quad \Delta Q_{\rm R}' & ~=~ q^*_1(i+\alpha) - Q(I) ~=~ \Delta Q' \big(\,1 - {\cal K}_1'\,\big),
  \end{split}
\end{equation}
where ${\cal K}_1' = f_3'(\nu,i)\, / \, \Delta V_1(I)$.  If we limit the slope of the interpolation profile to:
\begin{equation}
  \label{eq:1dpli_mono2}
  \Delta Q'' = 
  \begin{cases}
    \,\textrm{sign}\,\big(\Delta Q'\big)\, |\Delta Q_{\rm L}|\, / \, \big(1 + {\cal K}_1\big) & \textrm{if}~~ |\Delta Q_{\rm L}'| > |\Delta Q_{\rm L}| \\[6pt]
    \,\textrm{sign}\,\big(\Delta Q'\big)\, |\Delta Q_{\rm R}|\, / \, \big(1 - {\cal K}_1\big) & \textrm{if}~~ |\Delta Q_{\rm R}'| > |\Delta Q_{\rm R}| \\[6pt]
    \phantom{x}\Delta Q' & \textrm{otherwise},
  \end{cases}
\end{equation}
and, because of the properties of the PLI slope $\Delta Q'$, the first two cases of equation (\ref{eq:1dpli_mono2}) will never occur simultaneously.  Thus, the equation for conservative, monotonic, 1-D PLI in curvilinear coordinates is:
\begin{equation}
  \label{eq:1dpli_mono3}
  q^*_1(i+\alpha) ~=~ Q(I) + \Delta Q''\big( (2\zeta-1) - {\cal K}_1\big).
\end{equation}

The previous discussion applies to interpolation of momenta in the directions perpendicular to the component normal (\eg, $s_1$ in the 2- and 3-directions). The covariant procedure for linear interpolation in between coarse faces (\eg, $s_1$ in the 1-direction; equation \ref{eq:momlinint}) is as follows:
\begin{align}
  \label{eq:momlinint_curve}
  s_1(i+\alpha,j+\beta,k+\eta)\,\delta V_1(i+\alpha) ~=~ & (1 - \zeta)\, s_1(i,j+\beta,k+\eta)\,\delta V_1(i)\nonumber\\
  & + \zeta\, s_1(i+\nu,j+\beta,k+\eta)\,\delta V_1(i+\nu).
\end{align}

Finally, the generalised \citet{ll04} algorithm is adapted to curvilinear coordinates simply by replacing $\vec{B}$ (and $\vec{b}$) in equations (\ref{eq:lilistep1})--(\ref{eq:lililinear}) with the magnetic ``flux functions'':
\begin{equation}
  \label{eq:curvilili}
  \vec{\Psi} ~=~ \big( \Psi_1, \Psi_2, \Psi_3 \big) ~=~ \left( g_{21}\,g_{31}\,\frac{\Delta V_2}{\Delta x_2} B_1,\, g_{31}\,g_{32}\, B_2,\, g_{21}\, B_3 \right).
\end{equation}
With this modification, the solenoidal condition can be written in ``Cartesian-like'' form, regardless of the geometry:
\begin{equation*}
  \nabla\cdot\vec{\Psi} ~=~ \frac{\del \Psi_1}{\del x_1} + \frac{\del \Psi_2}{\del x_2} + \frac{\del \Psi_3}{\del x_3} ~=~ 0,
\end{equation*}
and the prolongation of magnetic field in curvilinear coordinates proceeds exactly as described in Section \ref{sub:spatialint}.
\section{The Vector Potential}
\label{app:vecpot}
Writing $\vec B = \nabla \times \vec A$ and assuming 3-symmetry with an appropriate gauge, one can easily show from the induction equation (equation \ref{eq:induction}) that $A_3$ obeys an ``advection equation":
\begin{equation}
\label{eq:dA3dt}
\frac{\partial A_3}{\partial t} + \vec v \cdot \nabla A_3 ~=~ 0,
\end{equation}
from which the poloidal components of the magnetic field are given by:
\begin{equation}
\label{eq:bpol}
B_1~=~ \frac{\partial A_3}{\partial x_2}; \qquad B_2 ~=\,\,-\frac{\partial A_3}{\partial x_1}.
\end{equation}
Further, one can show that $B_3$ is given by:
\begin{equation}
\label{eq:dB3dt}
\frac{\partial B_3}{\partial t} + \nabla \cdot (B_3 \vec v) ~=~ \vec B \cdot \nabla v_3.
\end{equation}
Since equation (\ref{eq:dB3dt}) has exactly the same form as the internal energy equation (equation \ref{eq:intnrg}), and since equation (\ref{eq:dA3dt}) is a simple advection equation, the induction step in the original \zeus code (\texttt{zeus04}; \citealt{c88}) was based on solving these equations using the hydrodynamical algorithms already in the code, and then calculating $B_1$ and $B_2$ from equations (\ref{eq:bpol}). Evidently, the solenoid condition is \emph{strictly} satisfied.  Regardless of the initial magnetic field configuration, it is easy to show that a face-centred $\vec B$ and a corner-centred (edge-centred in 3-D) $A_3$ guarantees $\nabla\cdot \vec B=0$ everywhere and at all times to machine round-off error.

However, \citet{c88} points out that a second-order accurate $A_3$ means first order accurate poloidal magnetic field components and a \emph{zeroth-order} accurate current density $J_3$ (each differentiation reduces the order of accuracy by one), and this was found to have catastrophic consequences in computing the $\vec J \times \vec B$ source terms in the momentum equation (equation \ref{eq:momentum}).  Thus, the vector potential algorithm in \texttt{zeus04} was abandoned, and the first publicly released version of \zeusdd (\citealt{sn92a,sn92b}), and later \zeusddd (\citealt{c96}), were based on the CT algorithm of \citet{eh88} in which the magnetic field is updated directly.  Note that the CT scheme \emph{conditionally} satisfies the solenoidal condition, requiring that the magnetic field be initialised such that $\nabla \cdot \vec B=0$.

Still, the failure of the vector potential algorithm in \texttt{zeus04} is not a complete indictment of $\vec A$ for use in numerical MHD algorithms.  Indeed, \citet{ldz00} and \citet{in02} have successfully demonstrated the use of $\vec A$ as the primary magnetic variable in their MHD codes.  By substituting $\vec B = \nabla \times \vec A$ into equation (\ref{eq:induction}) and with an appropriate gauge, we can write:
\begin{equation}
\label{eq:dAdt}
\frac{\partial \vec A}{\partial t} ~=~ \vec v\times \vec B ~\equiv\,\,- \vec E.
\end{equation}
Thus, CT can be used as originally designed in which $\nabla \times \vec E$ is used to update $\vec B$, or easily modified to use $\vec E$ to update $\vec A$ directly, then update $\vec B$ by taking a curl of the updated $\vec A$.  Either way, a curl must be taken and the algorithms are interchangeable to machine round-off error.  Based as they are on such a modified CT scheme, the vector-potential algorithms used by \citet{ldz00} and \citet{in02} are very different from the failed algorithm described for \texttt{zeus04}, showing none of the effects of the inaccurate current density.

Based on this observation, preliminary versions of \azeus used the vector potential as the primary magnetic variable so that prolongation could be accomplished by interpolating $\vec A$ (rather than $\vec B$), thus guaranteeing preservation of the solenoidal condition on a newly-created fine grid or in a fine boundary region.  Furthermore, because the vector potential conserves magnetic flux via a path integral ($\oint \vec{A}\cdot d\vec{l} = \int \vec{B}\cdot d\vec{\sigma}$), restricting $\vec A$ then calculating $\vec{B}$ means that no EMF corrections are required at adjacent boundaries.  This approach, however, was found to be unsatisfactory since the parabolic interpolation function generated by PPI on $\vec A$ produces piecewise-linear profiles for $\vec B$ and piecewise-constant profiles for $\vec J$, recovering the problem that doomed the original \texttt{zeus04} algorithm.

In addition, having to match gauges on overlapping grids poses a significant problem, and one that we never solved.  In order to arrive at equation (\ref{eq:dAdt}), one implicitly assumes a specific gauge.  For a single grid, this is not a problem since there is never a need to specify this gauge.  For multiple grids, however, each grid may have its own gauge (especially for grids whose origins are not coincident) and leaving them unspecified gives rise to discontinuities in the perpendicular components of the magnetic field at adjacent boundaries.  While a solution to this problem likely exists, we abandoned vector potentials in \azeus before finding one because of the insurmountable problem of the lack of accuracy in the current densities.

\begin{figure}[t!]
  \begin{center}
    \includegraphics[width=0.48\textwidth]{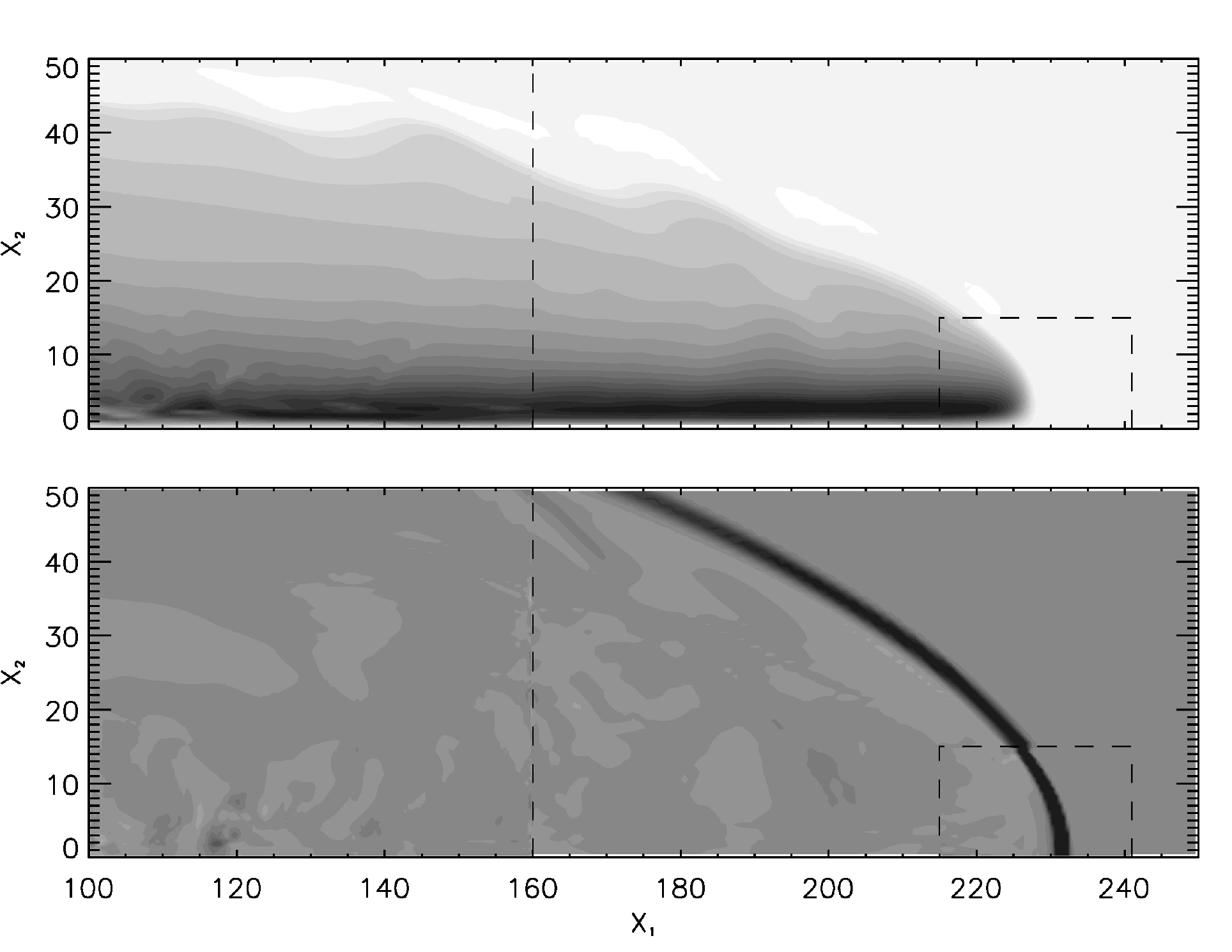}\hspace*{2.0ex}\includegraphics[width=0.48\textwidth]{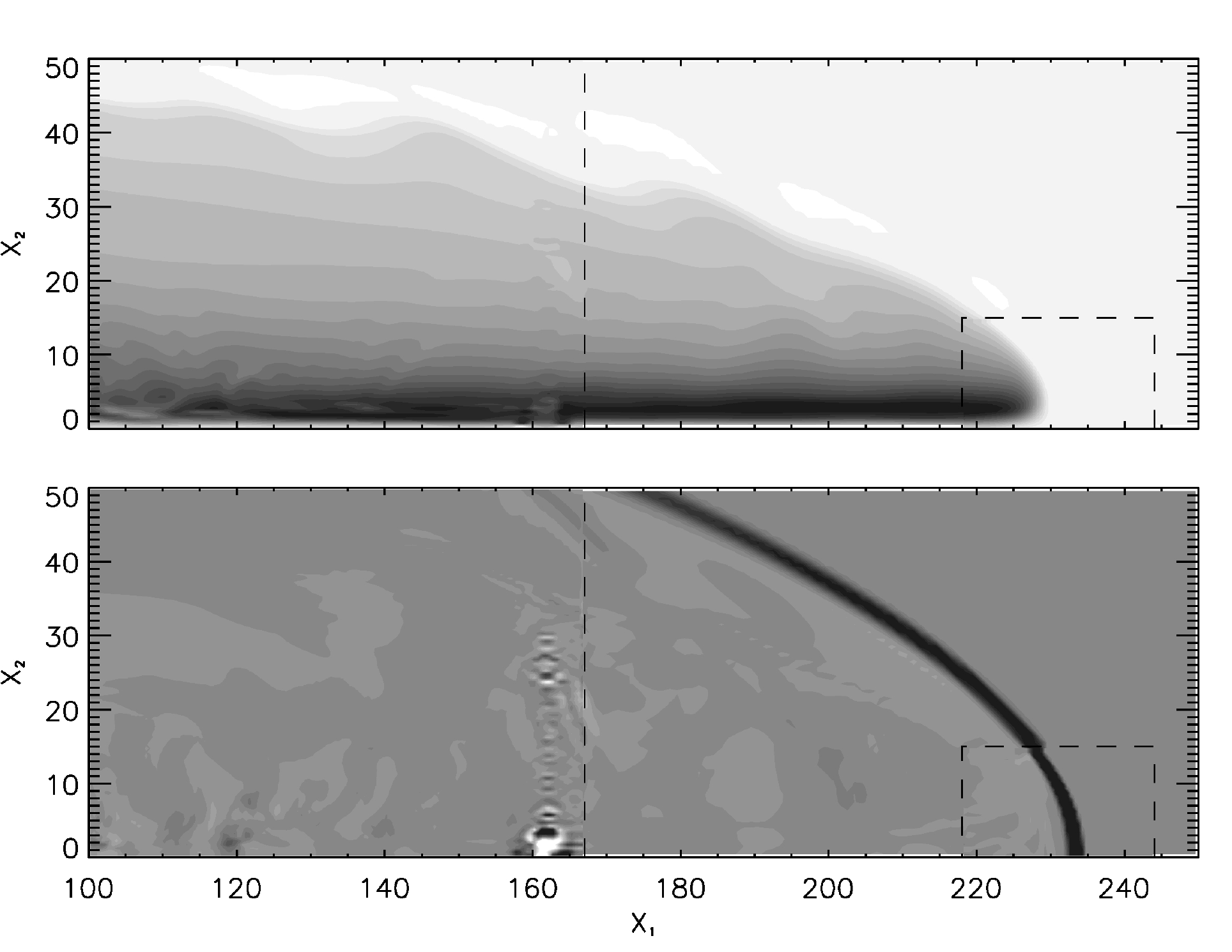}
  \end{center}
  \caption{\label{fig:vecpot} The effects of differencing a parabolic interpolation of $\vec{A}$ twice to calculate $\vec{J}\times\vec{B}$ forces.  Left panel:  the solution immediately before a grid adaptation step.  Right panel:  the solution a few time steps after.  Plotted are 20 evenly spaced contours of the toroidal magnetic field (top) and velocity divergence (bottom) with ranges $B_\varphi  = [-0.035, 0.0]$ and $\nabla\cdot\vec{v} = [-0.15, 0.15]$, respectively.}
\end{figure}

As an illustration, Figure \ref{fig:vecpot} shows early results of the simulations in \citet{rc11} in which the vector potential is used as the primary magnetic variable.  The left panel shows the solution immediately before a grid modification, and the right panel the solution a few time steps after and after the fine grid was extended from $x_1=160$ to $x_1=166$.  The errors committed by the piecewise constant current densities take a while to dissipate and, in this particular example, result in particularly egregious defects in the velocity divergence distribution within the new portion of the grid.  Conversely, the \citet{ll04} algorithm we currently employ renders adjacent boundaries virtually invisible in the distributions of all variables.

\section{The effects of a monopole in the boundary}
\label{app:monopole}

\begin{figure}[t!]
  \begin{center}
    \includegraphics[width=0.54\textwidth]{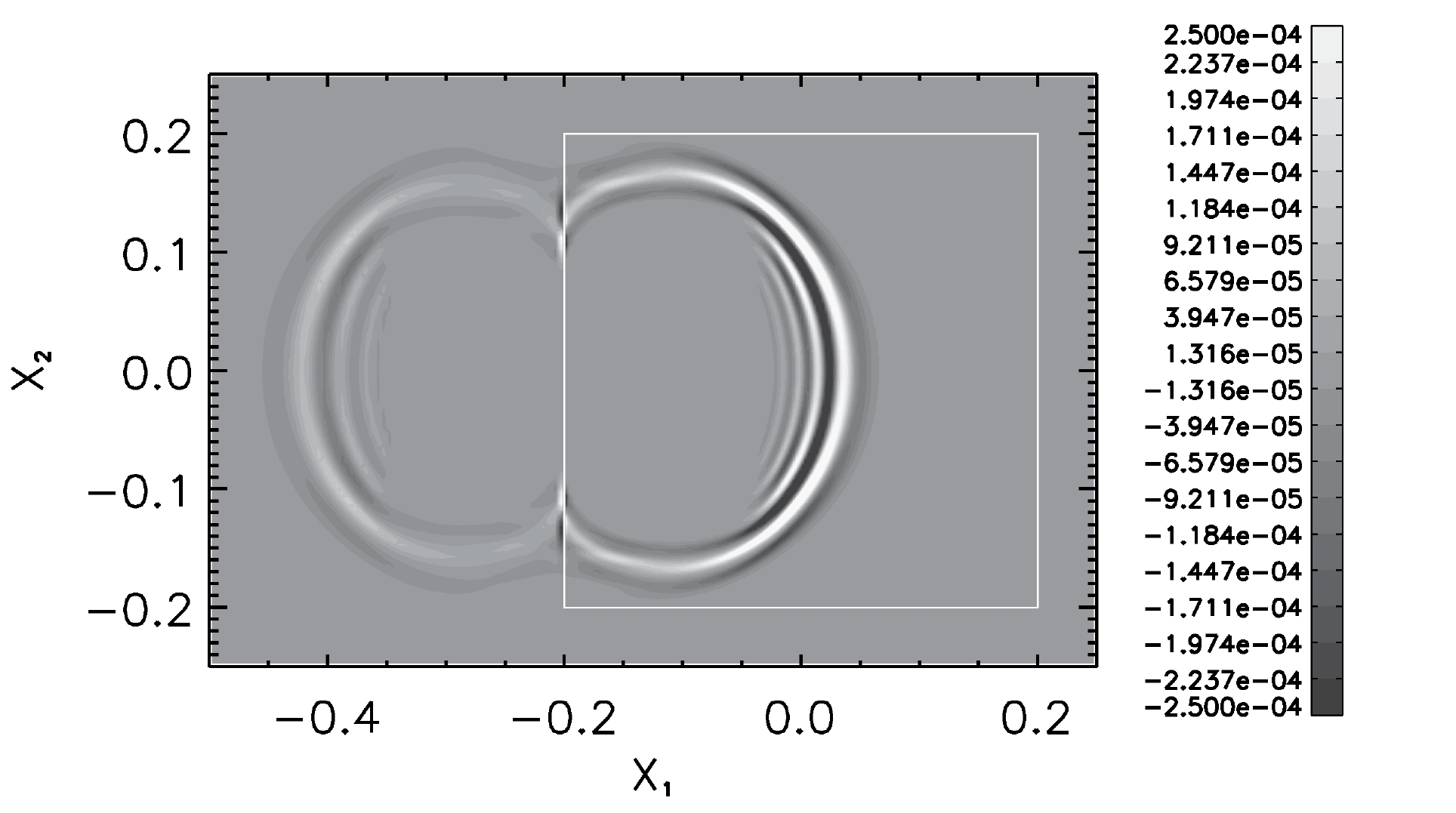}\hspace*{-3.75ex}\includegraphics[width=0.54\textwidth]{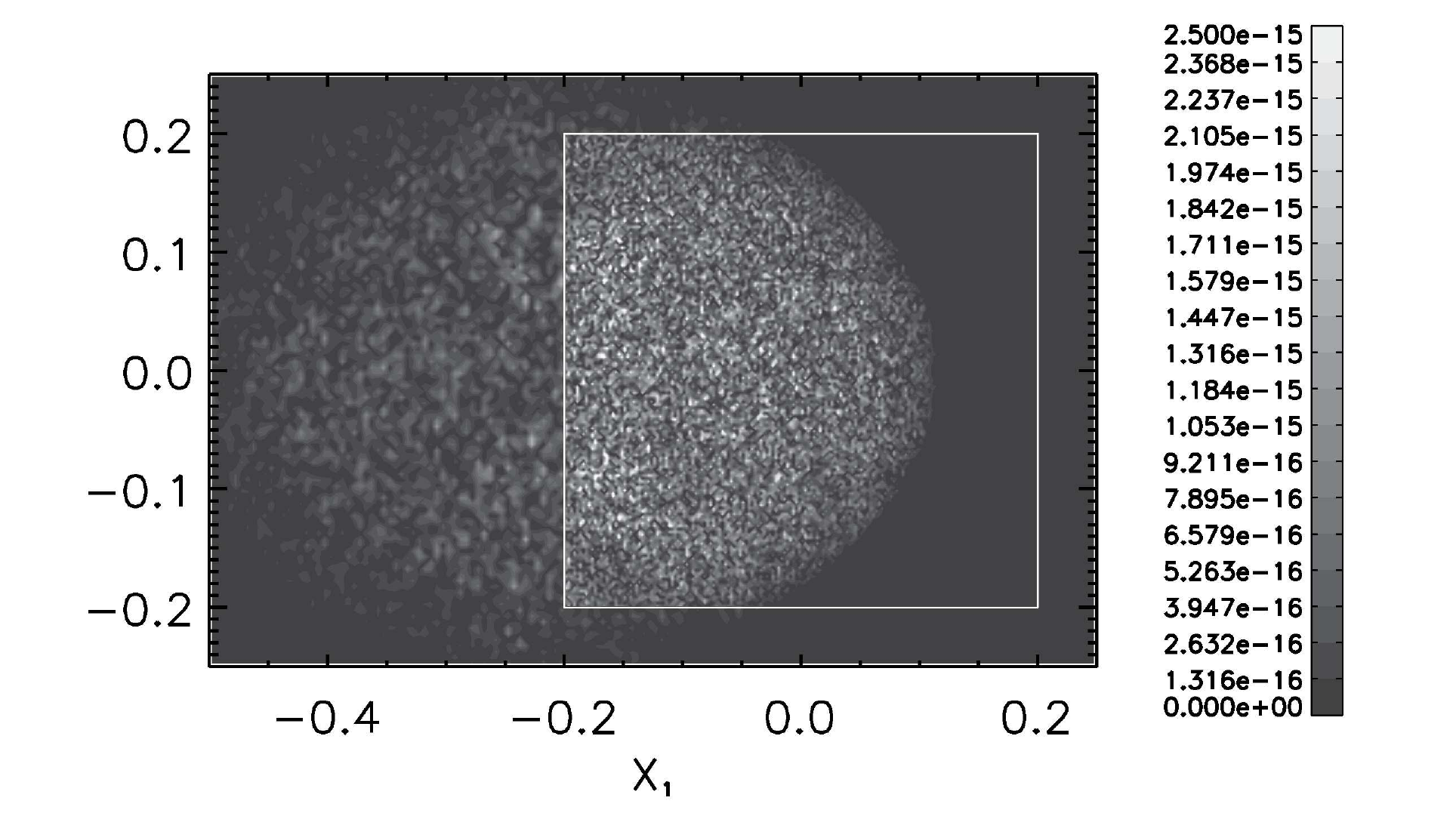}
  \end{center}
  \caption{\label{fig:monobdy} The results of placing a ``field defect'' at $x_2 = 0$ in the left boundary of a static grid for a single fine time step.  Plotted are contours of velocity divergence (left) and normalised magnetic field divergence (right).  The static grid boundaries are shown as a solid white line, and the simulation is depicted at $t = 0.17$.}
\end{figure}

Since \azeus is vulnerable to boundary pollution (Section \ref{sec:bcs}), we must take care that values in the self-computed boundaries of AMR grids are as free from non-physical phenomena (\eg, monopoles) as the grid itself.  To demonstrate this, Figure \ref{fig:monobdy} presents the results from a simulation with initially uniform and quiescent conditions $(\rho, p, \vec{v}, B_1, B_2, B_3) = (1, 0.6, 0, 0, 1, 0)$, where a static grid with corners $[\pm 0.2,\pm 0.2]$ is positioned in the centre of a Cartesian domain $(x_1,x_2) \in [-0.5,0.5]$.  A single ``field defect'' ($\vec B = 2B_{2}\hat x_2; \nabla\cdot\vec B\, \Delta x_2/B_2= \pm 1$) is deliberately placed in the left boundary of the static grid at $x_2 = 0$ and $t = 0$.  This field defect persists only for a single fine time step before the boundaries are replaced by data prolonged from the underlying coarse values in the manner described in Section \ref{sec:prolong}.  Other parameters used in this simulation include a refinement ratio of $\nu = 2$, $\gamma = 5/3$, ${\cal C} = 0.5$, \verb|qcon| = 1.0, and \verb|qlin| = 0.1 (as defined in Table \ref{tab:codeparam}).  The base grid has a resolution of $200 \times 200$ zones, and all physical boundaries are set to outflow conditions.

Although the field defect is only present for a single fine time step, a fast magnetosonic wave is still launched from the boundary of the fine grid, propagating into both fine and coarse grids.  All deviations from quiescence result from the momentary presence of the monopole in the boundary, leading to what is clearly an unacceptable result.  And yet, $\nabla \cdot \vec B = 0$ to within machine round-off error everywhere on the active grid at all times (\eg, Figure \ref{fig:monobdy}, right), and additionally within the boundaries after the first fine time step.  Therefore, the magnetic field prolongation and restriction operators in \azeus are designed to preserve magnetic divergence to machine round-off error in both the active grid and in the boundaries at all times.

\section{Schematic overview of the AMR module}
\label{app:amrmodule}
This appendix is designed mainly for programmers and those who may wish to use and/or modify \azeus.  It is meant as an overview to illustrate how the main ideas covered in this paper have been implemented in the code.

\begin{enumerate}[leftmargin=*,label=Step \arabic{*}.,start=0,align=left,labelsep=5pt]
\item \textbf{Initialise} computational domain and all variables for the current run.\\
\textbf{Set} level \verb|lvl = 1|, \verb|ntogo(lvl) = 1|.\\[2.0ex]
\textbf{MAIN LOOP:}
\item \textbf{For} level \verb|lvl|, \verb|call REGRID(lvl)| if \verb|kcheck| cycles have transpired at level \verb|lvl| since the last call to \verb|REGRID|, or if \verb|t = 0|.\\
\textbf{For} \verb|l = maxlevel - 1, lvl, -1:|
\begin{enumerate}[leftmargin=*,labelindent=\parindent,label=Step 1\alph{*}.,align=left, topsep=0pt,partopsep=0pt,labelsep=2pt]
  \item Flag existing grids at level \verb|l| for refinement based on one or more physical criteria.  Add \verb|ibuff| buffer zones around flagged points.
  \item Create grids around flagged points based on \verb|geffcy|.
  \item Check for proper nesting on all new/modified grids; fix grids which are not properly nested via bisection.
  \item If any new/modified grids are abutting, make them overlapping instead.
\end{enumerate}
\textbf{End For}\\
\textbf{For} \verb|l = lvl + 1, maxlevel:|
\begin{enumerate}[leftmargin=*,label=Step 1\alph{*}.,start=5,labelindent=\parindent, align=left,topsep=0pt,partopsep=0pt,labelsep=2pt]
  \item Fill new/modified grids with values either from old grids at level \verb|l| or interpolated from coarse grids at level \verb|l - 1|.
  \item Remove old grids which are no longer in use.
\end{enumerate}
\textbf{End For}
\item \textbf{For} level \verb|lvl|, \verb|call ADVANCE(lvl, dt)|.
\begin{enumerate}[leftmargin=*,label=Step 2\alph{*}.,labelindent=\parindent,align=left, topsep=0pt,partopsep=0pt,labelsep=2pt]
  \item \textbf{For} all grids at level \verb|lvl|, fill boundary zones either from overlapping grids at the same level, or interpolate values from coarse grids at level \verb|lvl - 1| (unless \verb|lvl = 1|, in which case use physical boundary values).
\end{enumerate}
\textbf{For} each grid \verb|m| at level \verb|lvl|:
\begin{enumerate}[leftmargin=*,label=Step 2\alph{*}.,start=2,labelindent=\parindent, align=left,topsep=0pt,partopsep=0pt,labelsep=2pt]
  \item Advance grid \verb|m| by time step \verb|dt| with \zeusddd.
  \item Save fluxes/EMFs along the edges of grid \verb|m| for flux corrections later.
\end{enumerate}
\textbf{End For}
\item One time step at level \verb|lvl| is complete; check for levels $>$ \verb|lvl|.\\
\textbf{Set} \verb|ntogo(lvl) = ntogo(lvl) - 1|.
\begin{enumerate}[leftmargin=*,label=Step 3\alph{*}.,labelindent=\parindent, align=left,topsep=0pt,partopsep=0pt,labelsep=2pt]
\item \textbf{If} \verb|lvl < maxlevel| then:
  \begin{itemize}[leftmargin=*,label=,labelindent=\parindent,align=left, topsep=0pt,partopsep=0pt,noitemsep,labelsep=2pt]
    \item \textbf{Set} \verb|lvl = lvl + 1|.
    \item \textbf{Set} \verb|ntogo(lvl) = nu|, where \verb|nu| is the refinement ratio.
    \item \textbf{Set} \verb|dt = dt / ntogo(lvl)|.
    \item \textbf{Go to} the beginning of the main loop.
  \end{itemize}
  \textbf{End If}
  \item \textbf{If} \verb|ntogo(lvl) > 0| then \textbf{go to} the top of the main loop,\\
  \textbf{Else Set} \verb|lvl = lvl - 1|.
\end{enumerate}
\item \textbf{For} level \verb|lvl|, \verb|call UPDATE(lvl)|.\\
\textbf{For} each grid \verb|m| at level \verb|lvl|:
\begin{enumerate}[leftmargin=*,label=Step 4\alph{*}.,labelindent=\parindent,align=left, topsep=0pt,partopsep=0pt,labelsep=2pt]
  \item Flux correct each grid \verb|m| with fluxes from grids at level \verb|lvl + 1|, if any.
  \item Overwrite zones on grid \verb|m| with overlying zones at level \verb|lvl + 1|, if any.
  \item Update any physical (non-periodic) boundary values which depend on zones which were just flux corrected or overwritten.
\end{enumerate}
\textbf{End For}\\
\textbf{If} \verb|lvl > 1|, go to Step 3b.
\item One entire AMR cycle is complete.  Reconcile time steps across all grids and levels.
\item Perform any required input/output.
\item \textbf{If} \verb|t < tlimit|, \textbf{go to} the top of the main loop, \textbf{else} exit.
\end{enumerate}
\end{document}